\documentclass[lettersize,journal]{IEEEtran}
\usepackage{bm} 

\usepackage{amsmath,amsfonts,amssymb}
\usepackage{mathtools} 
\usepackage{microtype} 
\usepackage{algorithmic}
\usepackage{algorithm}
\usepackage{array}
\usepackage[caption=false,font=footnotesize]{subfig}
\usepackage{textcomp}
\usepackage{stfloats}
\usepackage{url}
\usepackage{verbatim}
\usepackage{graphicx}
\usepackage{epstopdf}
\usepackage{cite}
\usepackage[table]{xcolor}
\usepackage{tikz}
\usetikzlibrary{positioning,arrows.meta,shadows,matrix}
\usepackage{enumitem}
\usetikzlibrary{arrows.meta,positioning}

\usepackage{booktabs}
\usepackage{multirow}

\newtheorem{theorem}{Theorem}

\newtheorem{definition}[theorem]{Definition}

\newtheorem{remark}[theorem]{Remark}

\newcommand{\R}{\mathbb{R}}
\newcommand{\Sphere}{\mathbb{S}}

\newcommand{\cV}{\mathcal{V}}            
\newcommand{\cE}{\mathcal{E}}            
\newcommand{\cM}{\mathcal{M}}            
\newcommand{\Mp}{\mathcal{M}_{+}}        

\newcommand{\KG}{\mathcal{G}_{\mathrm{KG}}}
\newcommand{\Ekg}{\mathcal{E}_{\mathrm{KG}}}
\newcommand{\NKG}[1]{\mathcal{N}_{\mathrm{KG}}(#1)}

\newcommand{\Tb}{\mathcal{T}_{b}}
\newcommand{\Pof}[1]{P(#1)}
\newcommand{\istar}{i^\star}

\newcommand{\vct}[1]{\bm{#1}}   
\newcommand{\mat}[1]{\bm{#1}}   
\newcommand{\vI}{\mat{I}}      

\newcommand{\vx}{\vct{x}}         
\newcommand{\vX}{\mat{X}}         
\newcommand{\vh}{\vct{h}}         
\newcommand{\vH}{\mat{H}}         
\newcommand{\vu}{\vct{u}}         
\newcommand{\vM}{\mat{M}}         

\newcommand{\vW}{\mat{W}}         
\newcommand{\vb}{\vct{b}}         

\newcommand{\vz}{\vct{z}}         

\begin{document}

\title{RieIF: Knowledge-Driven Riemannian Information Flow for Robust Spatio-Temporal Graph Signal Prediction in 6G Wireless Networks}

\author{Zhonghao~Jiu, Yongming~Huang,~\IEEEmembership{Fellow,~IEEE}, Fan~Meng, Hang~Zhan, Zening~Liu, Xiaohu~You,~\IEEEmembership{Fellow,~IEEE}
	\thanks{This work was supported in part by the National Natural Science Foundation of China under Grant No. 62225107, 62201394, the Natural Science Foundation on Frontier Leading Technology Basic Research Project of Jiangsu under Grant BK20222001, and the Fundamental Research Funds for the Central Universities under Grant 2242022k60002. (Corresponding authors: Y.\ Huang)}
	\thanks{Z. Jiu, Y. Huang and X. You are with the School of Information Science and Engineering, Southeast University, Nanjing 210096, China. Y. Huang, F. Meng, H. Zhan, Z. Liu and X. You are (also) with the Purple Mountain Laboratories, Nanjing 211111, China (e-mail: 230228222@seu.edu.cn; mengfan@pmlabs.com.cn; zhanxing@pmlabs.com.cn; huangym@seu.edu.cn; liuzening@pmlabs.com.cn; xhyu@seu.edu.cn).}
}



\maketitle

\begin{abstract}

With 6G evolving towards intelligent network autonomy, artificial intelligence (AI)-native operations are becoming pivotal. Wireless networks continuously generate rich and heterogeneous data, which inherently exhibits spatio-temporal graph structure. However, limited radio resources result in incomplete and noisy network measurements. This challenge is further intensified when a target variable and its strongest correlates are missing over contiguous intervals, forming \emph{systemic blind spots}. To tackle this issue, we propose RieIF (\emph{K}nowledge-driven \emph{R}iemannian \emph{I}nformation \emph{F}low), a geometry-consistent framework that incorporates knowledge graphs (KGs) for robust spatio-temporal graph signal prediction. For analytical tractability within the Fisher--Rao geometry, we project the input from a Riemannian manifold onto a positive unit hypersphere, where angular similarity is computationally efficient. This projection is implemented via a graph transformer, using the KG as a structural prior to constrain attention and generate a micro stream. Simultaneously, a Long Short-Term Memory (LSTM) model captures temporal dynamics to produce a macro stream. Finally, the micro stream (highlighting geometric shape) and the macro stream (emphasizing signal strength) are adaptively fused through a geometric gating mechanism for signal recovery. Experiments on three wireless datasets show consistent improvements under systemic blind spots, including up to 31\% reduction in root mean squared error and up to 3.2\,dB gain in recovery signal-to-noise ratio, while maintaining robustness to graph sparsity and measurement noise.

\end{abstract}


\section{Introduction}\label{sec:introduction}

As 6G visions move toward AI-native network operations, data-centric intelligent network autonomy is becoming increasingly important~\cite{zhou2022intelligence, 10183795, qin2024ai, 10522623}. Wireless networks continuously generate structured and coupled data across different layers, and the collected data can be constructed as a spatio-temporal graph. For example, Key Performance Indicators (KPIs) such as Signal-to-Interference-plus-Noise Ratio (SINR), Channel Quality Indicator (CQI), Modulation and Coding Scheme (MCS) reports, Hybrid Automatic Repeat Request (HARQ) feedback, and throughput are shaped by physical and protocol relations of fading, interference, scheduling, and link adaptation. However, observations of wireless data are often incomplete and noisy due to limited radio resources, imperfect hardware, and reporting pipelines. 

Data missingness is typically detrimental to network performance and stability~\cite{xie2019missingnetmeas}. When a data field that is strongly coupled with others becomes unavailable, multiple involved variables can be subsequently absent over contiguous time blocks, yielding structured missingness that violates independent and identically distributed (i.i.d.) assumptions in intelligence-driven operations. For instance, a missing observation of throughput can trigger the concurrent loss of cross-layer data fields, such as throughput, physical resource block (PRB) usage, transport-block size, and HARQ statistics. This regime is termed systemic blind spots. These blind spots do more than remove samples; they sever the correlation pathways that conventional fault propagation relies on, resulting in locally sparse evidence and ill-conditioned message passing. Consequently, robust recovery of spatio-temporal graph signals under systemic blind spots remains a meaningful and challenging problem. 

Knowledge-driven deep learning~\cite{sun2025comprehensive} integrates wireless domain knowledge into AI, and offers a novel perspective to address this challenge. As a structured form of knowledge representation, KGs~\cite{sun2025comprehensive} model entities along with their attributes and relations. By organizing information in graph form, KGs facilitate not only the efficient integration of heterogeneous data but also support rich relational reasoning. KG analysis enables the extraction of small but critical datasets suitable for lightweight AI models~\cite{10553365}, thereby promoting real-time intelligence~\cite{JFXG202501001}. Beyond data extraction, the semantically structured prior derived from a KG can also be embedded into AI model design to enhance tasks such as inference, completion, and robust prediction—especially when observations are partial or noisy. In this work, we focus on estimating missing wireless data fields and predicting their unobserved trajectories over spatio-temporal graphs. Within the broader graph signal processing literature, this task aligns closely with graph signal reconstruction. Particularly, the objective is to estimate only the unobserved entries of target data fields, rather than regenerating the entire graph signal. In practice, missing data may manifest in various forms, including sporadic gaps in observations of a subset of wireless data fields, contiguous blocks of complete missingness, or systemic blind spots where a target wireless data field and its strongest correlated proxies disappear simultaneously.

\subsection{Related Works}

We next review prior work along three lines: prediction in wireless communications, spatio-temporal recovery under missingness, and geometry-aware similarity for structured signal learning.

In wireless communications, prediction problems have been widely investigated to support proactive and intelligent operations across different layers. Existing approaches can be broadly categorized into knowledge-driven and data-driven paradigms. Knowledge-driven learning has been advocated for network optimization and maintenance~\cite{sun2025comprehensive}, while many graph neural network (GNN)-based methods are developed for reasoning and diagnosis, but works w.r.t. enhancement in models are rarely concerned. On the other hand, data-driven predictive tasks are deeply investigated across network layers: from network-level traffic forecasting via meta-learning~\cite{li2023meta}; to link-level beam prediction to reduce training overhead and delay in millimeter-wave (mmWave) systems~\cite{9269463, wang2025deep}; down to physical-layer channel estimation for vehicle-to-everything communications~\cite{zhang2024transformer}, unmanned aerial vehicle (UAV) links~\cite{chang2025beam}, non-terrestrial network (NTN) uplinks~\cite{minani2025channel}. Meanwhile, a pivotal aspect is often overlooked: the robustness of these predictors when critical data streams are missing—a frequent occurrence due to tight resource constraints.

When observations are incomplete, recovery methods exploit either global structure or graph-temporal dependencies. 1) Global low-rank approaches, such as Temporal Regularized Matrix Factorization (TRMF) and Bayesian tensor decompositions~\cite{yu2016temporal, chen2019bayesian}, can be effective when missingness is moderate and informative samples remain. 2) For graph-structured dynamics, spatio-temporal graph neural networks such as Spatio-Temporal Graph Convolutional Networks (STGCN), Graph WaveNet, and Attention Based Spatial-Temporal Graph Convolutional Networks (ASTGCN)~\cite{wu2019graph, guo2019attention}, as well as imputation-oriented variants including Graph Imputation Networks (GRIN) and SPatio-temporal Imputation Networks (SPIN)~\cite{cini2021filling, marisca2022learning}, learn dependency-driven propagation. 3) Generative completion, exemplified by Conditional Score-based Diffusion for Imputation (CSDI)~\cite{tashiro2021csdi}, and missingness-aware designs such as GinAR (graph information augmentation) and CoIFNet (collaborative information flow)~\cite{yu2024ginar, tang2025coifnet}, further enhance robustness through uncertainty modeling or collaborative information flow. These approaches typically rely on observable surrogates of the target, but systemic blind spots violate this premise because the target and its strongest proxies can disappear together.

To further improve robustness in structured learning, geometry-aware similarity has been explored. Information geometry endows statistical models with a Riemannian metric, where the Fisher--Rao metric yields invariant distances between distributions~\cite{amari2016information, cencov2000statistical}. Hyperspherical normalization and angular objectives emphasize direction over magnitude and can reduce sensitivity to scale variations~\cite{wang2020understanding}. Non-Euclidean graph representation learning further studies hyperbolic and curvature-based geometries~\cite{chami2019hyperbolic, ollivier2009ricci}. Prior works mainly use geometry as a representation choice for embedding or classification, rather than as the training criterion for spatio-temporal signal recovery under long correlated outages.

\begin{figure}[t]
    \centering
    \includegraphics[width=0.9\linewidth,trim=2mm 1mm 2mm 0mm,clip]{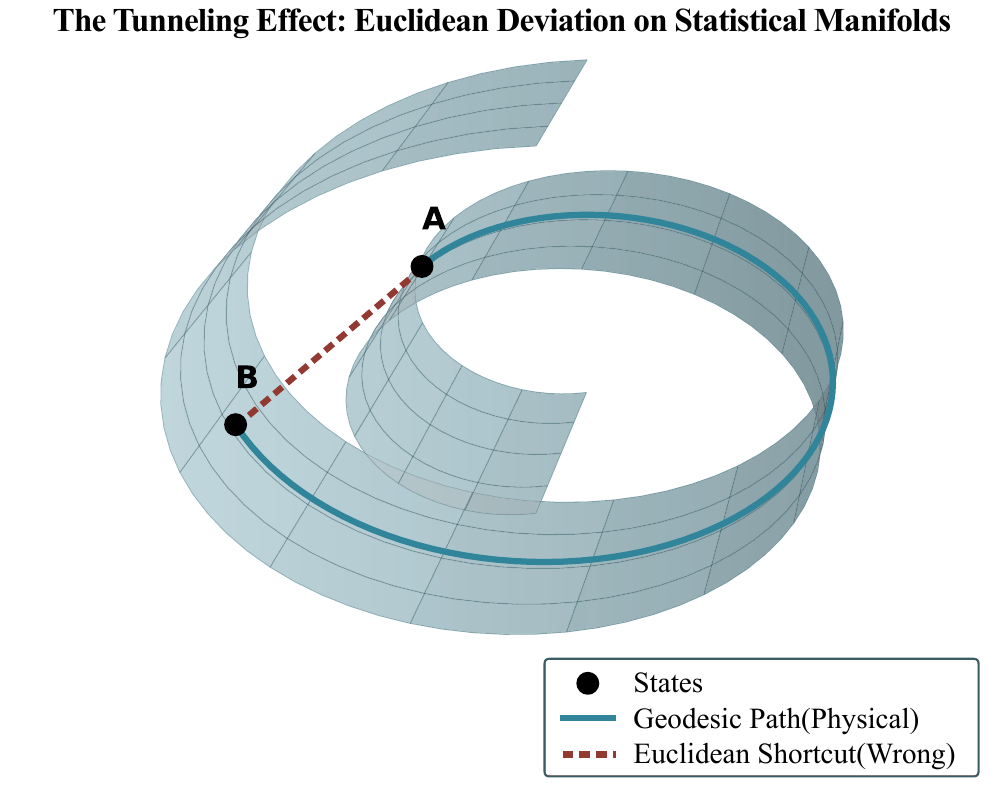} 
    \caption{Motivation: on a curved wireless state manifold, Euclidean recovery can tunnel across missing blocks, whereas geometry-consistent recovery follows a manifold geodesic.}
    \label{fig:euclidean_fallacy}
\end{figure}

In summary, systemic blind spots reveal a critical inductive-bias mismatch in conventional missing-data prediction pipelines, stemming from structured evidence removal and the curved manifold of network state evolution governed by coupled physical and protocol constraints. This mismatch manifests in three principal challenges:
\begin{itemize}
	\item \textbf{Collapsed information flow:} when a target and its strongest proxies are concurrently absent, correlation-based propagation becomes unreliable and learned dependency graphs become underconstrained.
	
	\item \textbf{Euclidean tunneling:} under sparse observations, Euclidean interpolation and dot-product aggregation may shortcut via infeasible chords across the manifold, deviating from physically feasible manifold evolution (Fig.~\ref{fig:euclidean_fallacy}).
	
	\item \textbf{Pronounced scale inconsistency:} the wide dynamic range of wireless data can cause amplitude-dominated similarity measures to obscure directional patterns shaped by interference, fading, and control loops.
\end{itemize}

\subsection{Our Contributions}

This paper investigates the robust recovery and short-horizon prediction of multivariate cross-layer wireless data, represented as spatio-temporal graph signals, under conditions of structured missingness. We specifically address the regime of systemic blind spots, where a target wireless data field and its most informative proxies are simultaneously absent over a contiguous interval. This regime is complementary to standard missing-entry settings and exposes a critical failure mode: when key statistical surrogates vanish, correlation-driven propagation collapses and Euclidean aggregation can shortcut through infeasible regions on a curved wireless state manifold. To maintain reliability under such conditions, we leverage a protocol-derived KG as a stable dependency backbone to compute correlations of structured data on a Fisher--Rao-consistent spherical chart, thereby mitigating Euclidean tunneling when observations are sparse and signal magnitudes are volatile.

To overcome this limitation, we propose RieIF, a knowledge-driven Riemannian information flow framework that formulates masked-index prediction as a geometry-consistent flow process rather than a Euclidean regression in $\mathbb{R}^n$. Our main contributions are summarized as follows.
\begin{itemize}
    \item \textbf{Systemic blind spots and reproducible evaluation protocol:} We formalize the notion of systemic blind spots for evolving wireless networks with structured and noisy measurements, defined as the simultaneous absence of a target wireless data field and its correlation-selected proxy set over a contiguous time block. Based on this formulation, we design a reproducible masking protocol for training and evaluation that deliberately induces structured evidence collapse.

    \item \textbf{Fisher--Rao-consistent geometry for robust prediction:} We map the intractable Fisher--Rao geodesic distance to computationally efficient angular similarity to capture geometric shape. Specifically, inputs are first projected into a positive domain via a smooth element-wise Softplus function, and then normalized onto a spherical chart. This two-step design reduces scale sensitivity and mitigates Euclidean tunneling under long missing blocks.
    
    \item \textbf{Knowledge-driven Riemannian information flow architecture:} We introduce a macro--micro dual-stream architecture that separately models the geometric shape and strength of missing data. The micro stream employs a time-invariant KG to guide geometry-aware, knowledge-constrained attention design in a graph transformer, while the macro stream uses an LSTM network to capture temporal dynamics. The two streams are adaptively fused via a positivity-preserving geometric gate, enabling stable information flow when data-driven correlations are unavailable.

    \item \textbf{Comprehensive validation under correlated outages:} Extensive experiments on three diverse wireless datasets, covering network-level throughput, system-level error vector magnitude (EVM), and link-level post-SINR predictions, demonstrate consistent performance gains under systemic blind spots. Achievements include up to 31\% reduction in root mean squared error and up to 3.2\,dB gain in recovery signal-to-noise ratio, with robustness to graph sparsity and measurement noise.
\end{itemize}

\section{System Model and Problem Formulation}
\label{sec:system_model}

This section introduces the observed spatio-temporal graph of wireless networks under systemic blind spots, and the protocol-derived knowledge graph served as an invariant structural prior. Then we formulate the masked-index recovery/prediction task of interest. Table~\ref{tab:notation} summarizes the main notations used throughout the paper.

\begin{table}[t]
\centering
\caption{Main Notation}
\label{tab:notation}
\renewcommand{\arraystretch}{1.12}
\setlength{\tabcolsep}{6pt}
\begin{tabular}{p{0.30\columnwidth} p{0.62\columnwidth}}
\toprule
\textbf{Symbol} & \textbf{Meaning} \\
\midrule
$\cV=\{1,\dots,N\}$ & Index set of wireless data/state variables (nodes); $N$ is the number of variables. \\
$\mathcal{T}=\{1,\dots,T\}$ & Discrete time indices; $T$ is the segment length. \\
$\Tb$ & Contiguous time block corresponding to a systemic blind spot. \\
$\KG=(\cV,\Ekg)$ & Protocol-derived knowledge graph; $\Ekg$ is its directed edge set. \\
$\NKG{i}$ & Knowledge-graph incoming neighborhood of node $i$: $\{j\mid (j,i)\in\Ekg\}$. \\

$\mat{Y}_{1:T}\in\R^{N\times T}$ & Standardized data-field trajectory with entries $\tilde{x}_{i,t}$. \\
$\vM_{1:T}\in\{0,1\}^{N\times T}$ & Missingness mask (1 indicates missing). \\
$\istar\in\cV$ & Target wireless data-field node in the systemic blind-spot protocol. \\
$\Omega_{\mathrm{tar}}$ & Target index set for recovery/evaluation, typically $\{(i,t)\mid \vM_{i,t}=1\}$. \\
$\widehat{\mat{Y}}_{1:T}=[\hat{x}_{i,t}]$ & Estimator output; $\hat{x}_{i,t}$ estimates $\tilde{x}_{i,t}$. \\
$\Pof{\istar}$ & Proxy set for the target node $\istar$ in the blind spot definition. \\

$K,\tau$ & Time-delay embedding dimension and delay. \\
$x^{\mathrm{raw}}_{i,t},\ \tilde{x}_{i,t}$ & Raw observation and its standardized (z-score) version for node $i$ at time $t$. \\
$\vx_{i,t}\in\R^{K}$ & Time-delay embedding (phase-space vector) of node $i$ at time $t$. \\
$\vX_t\in\R^{N\times K},\ \vX_{1:T}$ & Phase-space snapshot stacking all $\vx_{i,t}$ at time $t$; sequence over $t=1,\dots,T$. \\

$D$ & Latent dimension in the positive-cone representation. \\
$\Phi_{\text{node}}(\cdot)$ & Node-wise lifting map into the positive latent space $\Mp\subseteq\R^{D}_{+}$. \\
$\vh_{i,t}\in\Mp$ & Latent positive state for node $i$ at time $t$. \\
$\vz_{i,t}\in\Sphere^{D-1}_{+}$ & Spherical representative: $\vz_{i,t}=\vh_{i,t}/\|\vh_{i,t}\|_2$. \\
$\vu^{\text{macro}}_{i,t},\ \vu^{\text{micro}}_{i,t}$ & Macro and micro tangent updates. \\
$\vct{g}_{i,t}\in(0,1)^{D}$ & Gate for fusing micro and macro updates. \\
\bottomrule
\end{tabular}
\end{table}

\subsection{System Model}
\label{subsec:obs_model}

We model wireless data measurements as a spatio-temporal graph, where each node corresponds to a protocol-aligned wireless data field and time indexes the observation sequence. Specifically, the node set is denoted as $\cV=\{1,\dots,N\}$ and the counterpart measurements are sampled at discrete time steps $t\in \mathcal{T} = \{1,\dots,T\}$. Let $x^{\mathrm{raw}}_{i,t}\in\R$ denote the raw measurement of node $i$ at time $t$, and $\tilde{x}_{i,t}$ be its standardized (z-score) version. Stacking all variables yields the standardized spatio-temporal graph signal
\begin{equation}
\mat{Y}_{1:T} = [\tilde{x}_{i,t}]_{i\in\cV,\ t\in\mathcal{T}} \in \R^{N\times T}.
\end{equation}
Missing entries in $ \mat{Y}_{1:T} $ are indicated by a binary mask $\vM_{1:T}\in\{0,1\}^{N\times T}$, where $\vM_{i,t}=1$ signifies that $x^{\mathrm{raw}}_{i,t}$ (and thus $\tilde{x}_{i,t}$) is unavailable, and $\vM_{i,t}=0$ otherwise. Throughout, we use the term prediction to cover both missing-entry recovery within the segment and short-horizon forecasting; the latter can be modeled by masking future time indices.

\subsection{Problem Formulation}
\label{sec:problem_formulation}

Given the partially observed spatio-temporal graph signal $\mat{Y}_{1:T}$ with its mask $\vM_{1:T}$, our goal is to estimate the unobserved entries in a target index set $\Omega_{\mathrm{tar}}\subseteq \cV\times\mathcal{T}$, typically chosen as the missing indices $\Omega_{\mathrm{tar}}=\{(i,t)\mid \vM_{i,t}=1\}$. Let $f_\theta$ be a parametric estimator that maps the available observations (and optional prior information) to a complete estimate:
\begin{equation}\label{eq:model_mapping}
\widehat{\mat{Y}}_{1:T} = f_\theta(\mat{Y}_{1:T},\, \vM_{1:T},\, \mathcal{G}),
\end{equation}
where $\widehat{\mat{Y}}_{1:T}=[\hat{x}_{i,t}]$ and $\mathcal{G}$ denotes an optional structural prior (e.g., a protocol-derived knowledge graph detailed introduced later in subsection~\ref{sec:kg_topology}). The learnable parameter $\theta$ is optimized to minimize the expected mean squared error on $\Omega_{\mathrm{tar}}$:
\begin{equation}\label{eq:opt_target}
\theta^\star = \operatorname*{argmin}_{\theta}\,
\mathbb{E}_{(\mat{Y}_{1:T},\vM_{1:T}) \sim \mathcal{D}}
\bigg[
\frac{1}{|\Omega_{\mathrm{tar}}|}
\sum_{(i,t)\in \Omega_{\mathrm{tar}}}
(\hat{x}_{i,t}- \tilde{x}_{i,t})^2
\bigg],
\end{equation}
where $\mathcal{D}$ is induced by sampling training segments and generating masks according to the missingness protocol.


{\bf Systemic blind spots.}
We focus on structured outages where a target wireless data field and its strongest proxies are simultaneously missing over a contiguous time block, which removes both the target and its most informative surrogates from the observations.

\begin{definition}[Systemic Blind Spot (Target + Proxies + Block Missingness)]
	\label{def:blind_spot}
	A systemic blind spot for a target node $\istar$ occurs over a contiguous time block $\Tb\subseteq\mathcal{T}$ when $\vM_{\istar,t}=1$ and $\vM_{j,t}=1$ for all $j\in \Pof{\istar}$ and all $t\in\Tb$. In our experiments, $\Pof{\istar}$ is instantiated by a correlation-threshold rule (cf.\ the masking protocol), but the definition is agnostic to how proxies are selected.
\end{definition}
This definition subsumes standard block missingness (by setting $|\Pof{\istar}|=0$) and sporadic proxy-target dropout (by setting $|\Tb|=1$), while the general case stresses cross-variable reasoning under correlated outages. This dependency collapse is particularly challenging: when the target and its strongest proxies are concurrently absent, correlation-driven propagation becomes underconstrained and recovery must rely on weaker evidence. Systemic blind spots can compromise purely data-driven dependency graphs, as a target wireless data field and its strongest proxies may vanish concurrently. Conversely, a time-invariant structural prior informed by wireless domain knowledge can address this shortcoming.

\subsection{Protocol-Derived Knowledge Graph}\label{sec:kg_topology}

\begin{figure}[t]
	\centering
	\includegraphics[width=1\linewidth,trim=2mm 2mm 2mm 2mm,clip]{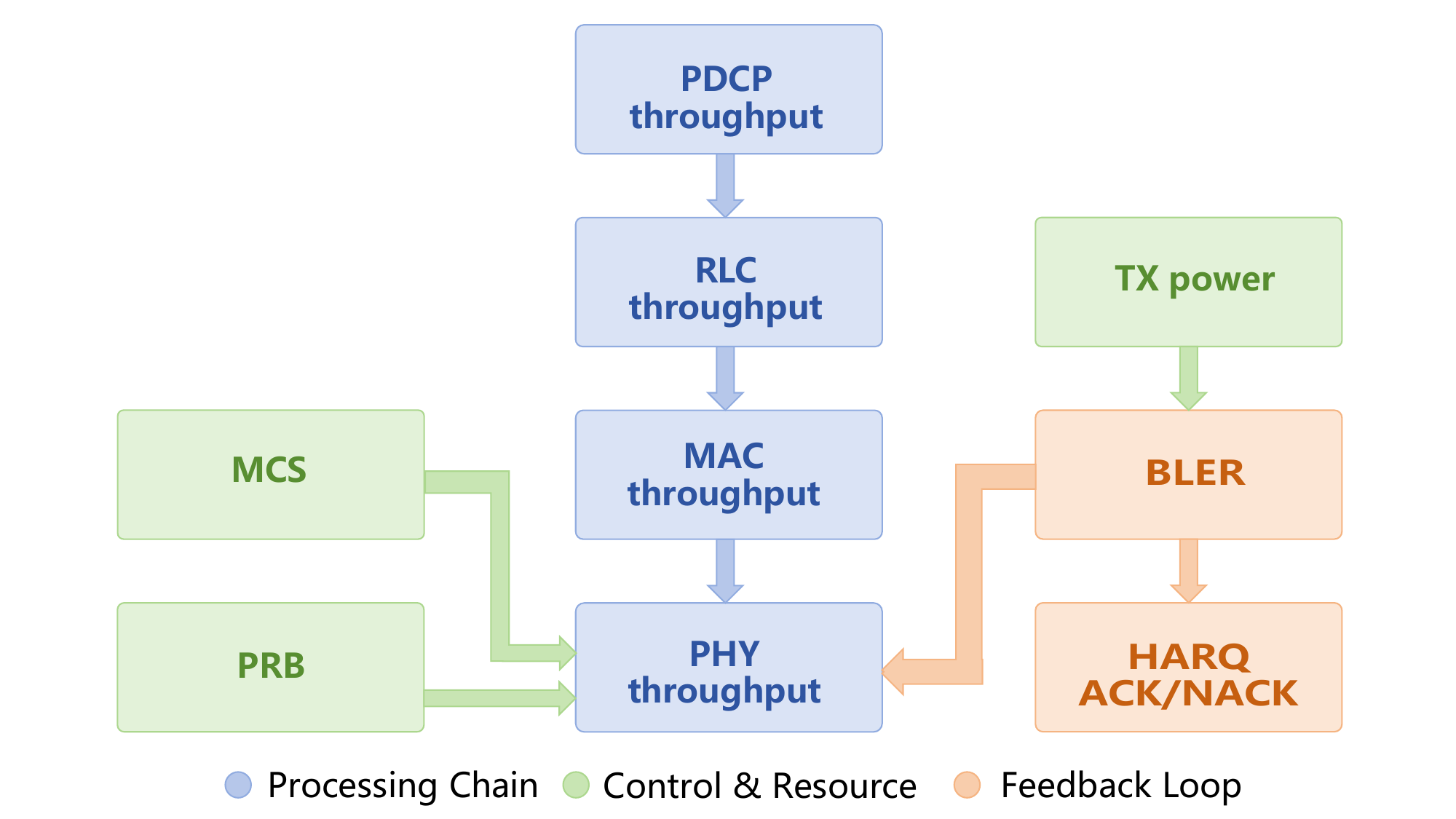} 
	\caption{Simplified protocol-derived knowledge graph showing vertical (cross-layer) and horizontal (intra-layer) dependencies among protocol-aligned wireless data fields.}
	\label{fig:kg_vertical}
\end{figure}

We construct a time-invariant protocol-derived knowledge graph $\KG=(\cV,\Ekg)$ that encodes cross-layer data-field dependencies grounded in 3GPP procedures and basic physical causality~\cite{3gpp_38_300,3gpp_38_214}. Unlike time-varying physical connectivity, the wireless data KG captures logical/protocol dependencies and thus remains as an invariant structural prior even when radio links or user locations change.

Edges are oriented from conditioning wireless data fields to the data fields they influence. For each node $i$, we define the incoming KG neighborhood as $\NKG{i}:=\{j\mid (j,i)\in\Ekg\}$, and we aggregate information along these directed dependencies. At a high level, $\Ekg$ contains (i) vertical edges that follow cross-layer processing order and (ii) horizontal edges that capture intra-layer control loops. We set $\Ekg=\mathcal{E}_{\text{vert}}\cup\mathcal{E}_{\text{hori}}$. In the following, we propose to utilize $\KG$ as an invariant structural prior to constrain information flow during prediction, when correlation routes collapse under systemic blind spots.

Fig.~\ref{fig:kg_vertical} illustrates a representative subgraph, and we summarize node semantics and deterministic edge-construction rules used to instantiate $\Ekg$.

\noindent\textbf{Node semantics:}
Nodes are selected from uplink-relevant groups: (i) modulation and coding indicators, including the MCS index and modulation-order ratios; (ii) throughput across the Packet Data Convergence Protocol (PDCP), Radio Link Control (RLC), Medium Access Control (MAC), and physical layer (PHY) processing chain; (iii) channel quality and reliability metrics, such as block error rate (BLER) and path loss; (iv) spatial multiplexing statistics, such as rank indicator (RI); and (v) resource and power control metrics, including PRB allocation, transmit power (TxP), and power headroom.

\noindent\textbf{Edge construction rules:}
The edge set $\Ekg$ follows deterministic protocol rules:
\begin{itemize}[leftmargin=*, nosep] 
    \item \textbf{Vertical edges} ($\mathcal{E}_{\text{vert}}$): Follow the stack processing order $\text{PDCP} \to \text{RLC} \to \text{MAC} \to \text{PHY}$, including dual-connectivity counterparts.

    \item \textbf{Horizontal edges} ($\mathcal{E}_{\text{hori}}$): Capture intra-layer control loops, specifically:
    (i) resource allocation ($\{\text{MCS}, \text{PRB}, \text{RI}\} \to \text{PHY}_{\text{thpt}}$); 
    (ii) power control ($\text{Pathloss} \to \text{TxP} \to \text{BLER}$); 
    and (iii) feedback loops ($\text{BLER} \to \text{ACK/NACK}$), where ACK/NACK denotes acknowledgment/negative acknowledgment.
\end{itemize}

\section{Proposed Method: The RieIF Architecture}\label{sec:proposed_method}

Building upon the system model and problem formulation, we begin by introducing geometric and attention primitives that leverage the information geometry. Subsequently, we develop the complete RieIF architecture and describe its detailed design.

\subsection{Preliminaries}\label{sec:preliminaries}

This subsection collects the geometric and attention primitives used by RieIF. These definitions are independent of the proposed macro--micro architecture and are stated up front to streamline the subsequent method description.

\noindent\textbf{Motivation: Manifold Constraints and Euclidean Tunneling:}

Let $\cM\subseteq \R^{D}_{+}$ denote the (unknown) feasible set of latent states induced by coupled physical and protocol relations, including rate adaptation loops, scheduling and queueing constraints, cross-layer processing, and interference coupling.
As a result, feasible trajectories are typically curved and nonconvex in the ambient space.

When a contiguous block of observations is missing, many learning-based predictors effectively interpolate in Euclidean space. For a curved manifold, the Euclidean chord $(1-s)\vh_{a}+s\vh_{b}$ for $s\in[0,1]$ can cut through infeasible regions,
which may produce physically infeasible states, as illustrated by the tunneling effect in Fig.~\ref{fig:euclidean_fallacy}. This geometric misalignment motivates the need for geometry-consistent aggregation that respects the intrinsic structure of $\cM$.

\noindent\textbf{Fisher--Rao-Aligned Spherical Chart:}

Direct geodesic computation on $\cM$ is intractable. Instead, we adopt a representation-induced chart that (i) preserves positivity via Softplus (cf. \eqref{eq:softplus}) and (ii) yields a computable surrogate distance aligned with information geometry.

Spherical chart and Fisher--Rao consistency:
To emphasize directional patterns over magnitude, we normalize latent vectors onto the positive unit hypersphere:
\begin{equation}
\vz_{i,t} \;=\; \frac{\vh_{i,t}}{\|\vh_{i,t}\|_2} \in \Sphere^{D-1}_{+}.
\label{eq:spherical_mapping}
\end{equation}
This disentanglement is physically meaningful for cross-layer wireless measurements in wireless networks: large-scale fading and power control mainly affect magnitudes, whereas interference coupling and multipath effects manifest as pattern (direction) changes.
To connect this spherical chart to information geometry, we convert the normalized direction into a probability object.
Define $\vct{p}=\vz\odot\vz\in\Delta^{D-1}$ (the probability simplex), induced by the positive-cone representation (no probabilistic assumption on raw wireless measurements).
Note that $\vct{p}$ is a valid simplex point because $\sum_{d=1}^{D} p_d = \sum_{d=1}^{D} z_d^{2} = \|\vz\|_2^{2} = 1$, and $p_d>0$ for all $d$ because Softplus yields strictly positive coordinates.
Under the square-root simplex chart, Fisher--Rao distances reduce to spherical angles, yielding the closed-form relation below.

With the standard convention in information geometry, the Fisher--Rao geodesic distance equals twice the spherical angle~\cite{amari2016information,cencov2000statistical}:
\begin{equation}
d_{\mathrm{FR}}(\vct{p},\vct{q}) \;=\; 2\arccos\!\left( (\vct{p}^{1/2})^{\top}(\vct{q}^{1/2}) \right).
\end{equation}
With $(\vct{p}^{1/2})=\vz$, this yields
\begin{equation}
d_{\mathrm{FR}}(\vct{p}_{i,t},\vct{p}_{j,t}) \;=\; 2\arccos\!\left(\vz_{i,t}^{\top}\vz_{j,t}\right),
\label{eq:fisher_rao_dist}
\end{equation}
which motivates geometry-aware aggregation based on angular similarity on $\Sphere^{D-1}_{+}$.

\begin{remark}[Representation-induced Fisher--Rao chart]
	\label{rem:induced_fr}
	The simplex vector $\vct{p}=\vz\odot\vz$ is induced by the learned positive coordinates and is not assumed for raw wireless measurements.
	We therefore use Fisher--Rao geometry as a shape-aware similarity in the learned chart.
	In implementation, cosine similarity $\vz_{i,t}^{\top}\vz_{j,t}$ is used as a monotone proxy of~\eqref{eq:fisher_rao_dist}, avoiding explicit evaluation of $\arccos(\cdot)$.
\end{remark}

\noindent\textbf{Interpretation:}
Wireless measurements are typically reported as windowed statistics of underlying random link and traffic processes. Over short intervals with approximately stationary operating points, the induced chart supports shape--scale disentanglement: angular similarity captures directional patterns while $\|\vh_{i,t}\|_2$ retains magnitude.

\begin{figure}[t]
	\centering
	\includegraphics[width=0.95\linewidth,trim=2mm 2mm 2mm 0mm,clip]{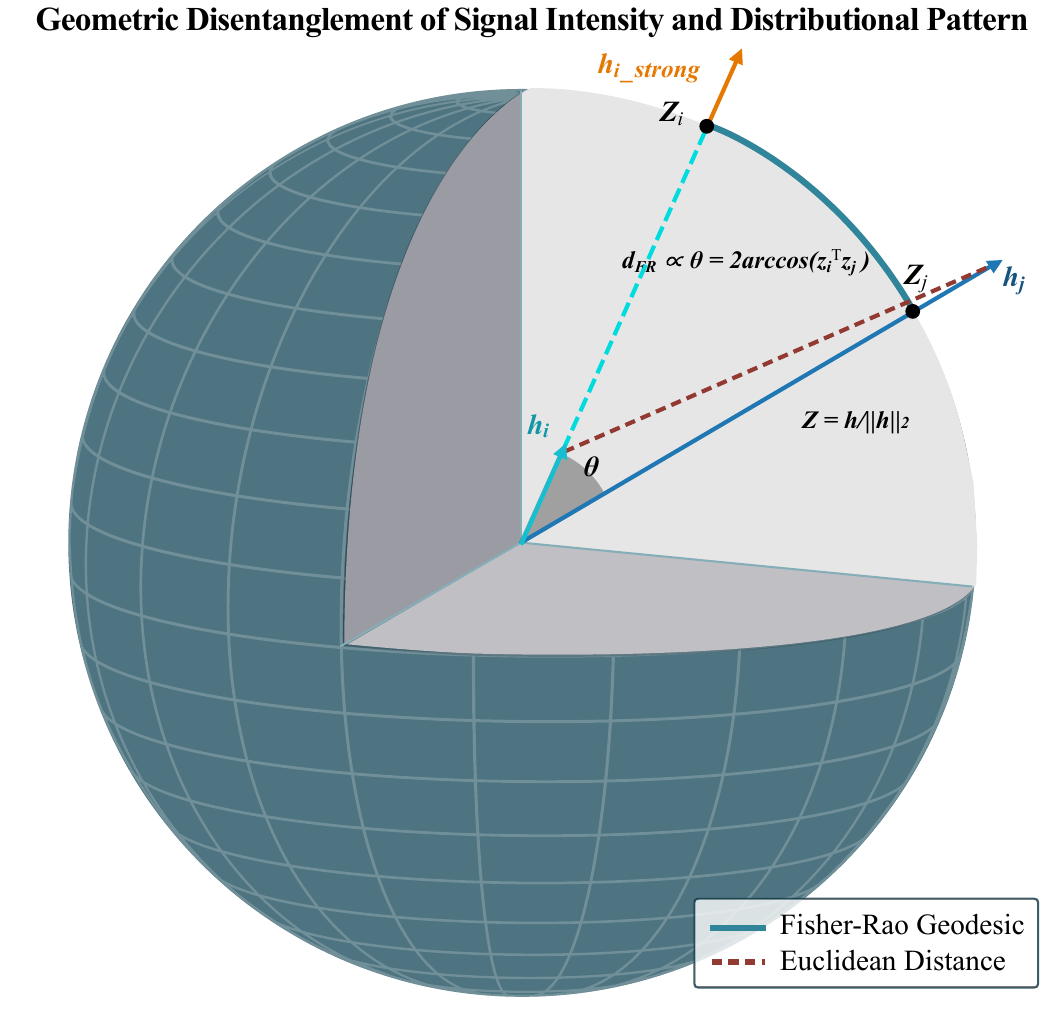}
	\caption{Spherical normalization disentangles shape (direction) from intensity (norm): vectors with the same direction map to the same point on $\Sphere^{D-1}_{+}$. RieIF matches shapes via angular similarity while modeling magnitude separately.}
	\label{fig:geo_disentangle}
\end{figure}

\noindent This shape--scale disentanglement motivates the hybrid training objective in Sec.~\ref{sec:synthesis}, which matches both magnitude (scale) and directional trends (shape).

\noindent\textbf{Scaled Dot-Product Attention and Laplacian Positional Encoding:}

We follow the standard scaled dot-product attention used in Transformers~\cite{vaswani2017attention} and graph attention variants~\cite{velickovic2017graph}; RieIF later specializes this primitive with a protocol-consistent hard mask to enforce KG-constrained information flow (Sec.~\ref{sec:spatial_stream}).

Laplacian positional encoding: Besides the instantaneous states, we inject a topology-aware positional encoding derived from the knowledge-graph Laplacian. Let $\mat{A}\in\{0,1\}^{N\times N}$ be the (symmetrized) adjacency of $\KG$ and let $\mat{L}=\vI-\mat{D}^{-1/2}\mat{A}\mat{D}^{-1/2}$ be the normalized Laplacian, where $\mat{D}$ is the degree matrix. We take the $d_{\mathrm{pe}}$ smallest nontrivial eigenvectors of $\mat{L}$ and denote the encoding for node $i$ by $\vct{e}_i\in\R^{d_{\mathrm{pe}}}$. We inject $\vct{e}_i$ into the attention projections:
\begin{equation}
\vct{q}_{i,t}=\vz_{i,t}\vW_Q+\vct{e}_i\vW_Q^{\mathrm{pe}},\qquad
\vct{k}_{j,t}=\vz_{j,t}\vW_K+\vct{e}_j\vW_K^{\mathrm{pe}},
\label{eq:lap_pe}
\end{equation}
\noindent where $\vW_Q^{\mathrm{pe}},\vW_K^{\mathrm{pe}}\in\R^{d_{\mathrm{pe}}\times d_k}$ are learnable.

\subsection{Architecture Overview and Macro--Micro Information Flow}
\label{sec:dynamics}

\begin{figure*}[!t]
   \centering
   \includegraphics[width=1\textwidth,trim=2mm 2mm 2mm 2mm,clip]{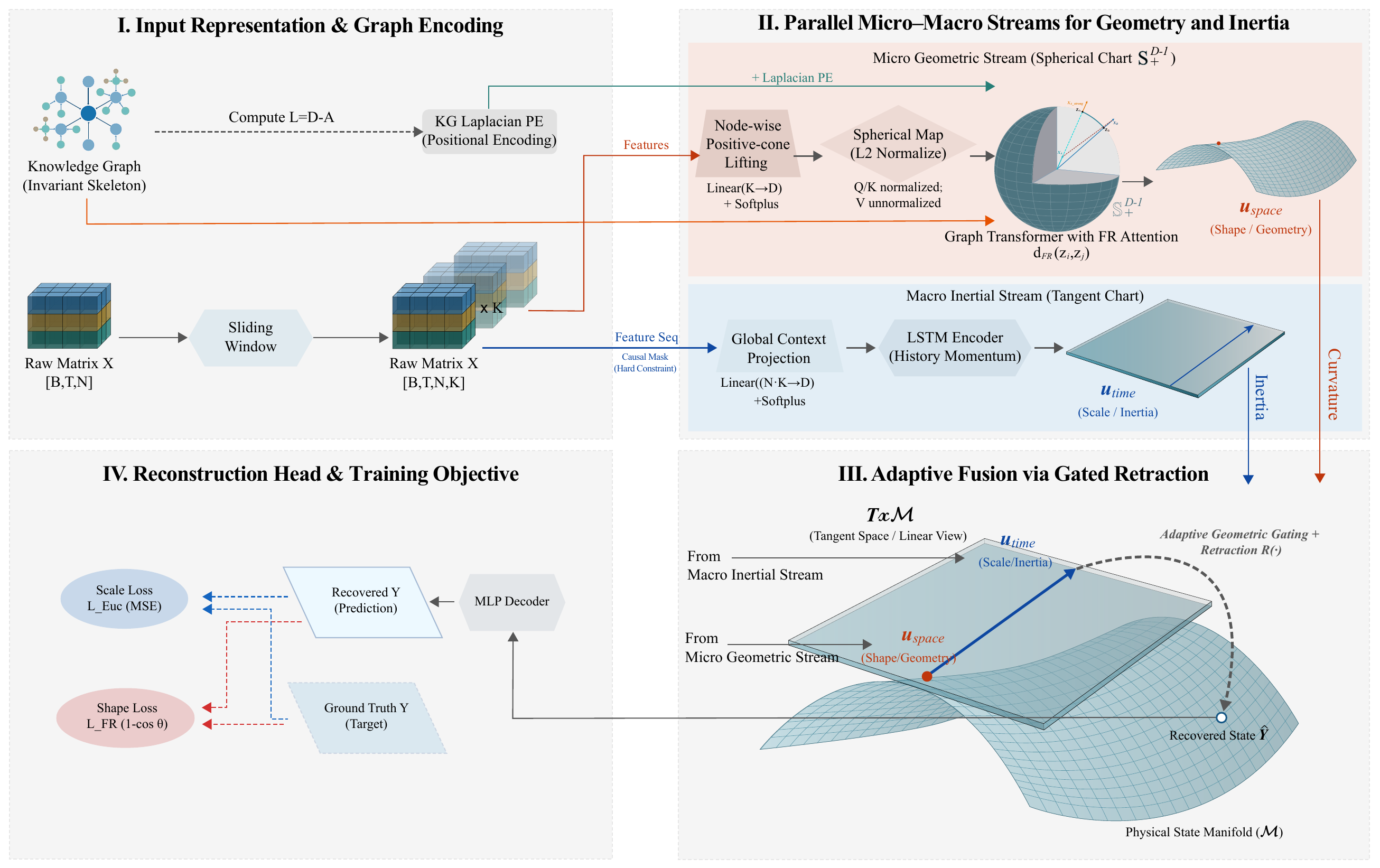}
   \caption{RieIF architecture. Phase-space inputs are embedded and lifted to a positive latent cone (Sec.~\ref{sec:signal_rep}); the macro stream models temporal inertia (Sec.~\ref{sec:temporal_stream}), and the micro stream performs Fisher--Rao-aligned attention on the spherical chart (Sec.~\ref{sec:preliminaries}) under the protocol-derived knowledge graph (Sec.~\ref{sec:kg_topology}). The two tangent updates are fused by a geometric gate and retracted for robust prediction.}
   \label{fig:framework}
\end{figure*}

As illustrated in Fig.~\ref{fig:framework}, RieIF adopts a macro--micro dual-stream design on a positivity-preserving latent chart. The macro stream captures network-level inertia and provides a stable fallback under systemic blind spots, while the micro stream performs geometry-aware aggregation along the protocol-derived knowledge graph $\KG$ (Sec.~\ref{sec:kg_topology}). To mitigate amplitude sensitivity, the micro stream computes similarity on a spherical chart, whereas the actual updates are accumulated in a Euclidean (tangent) chart and then retracted to the positive cone via a smooth positivity-preserving map.

\noindent\textbf{Computational workflow:}
At each time step $t$, we form the masked phase-space snapshot $\vX_t\in\R^{N\times K}$ via time-delay embedding in~\eqref{eq:phase_embed} (Sec.~\ref{sec:signal_rep}), where unobserved samples are zero-filled according to $\vM_{1:T}$. Node-wise embeddings are lifted to $\vH^{(0)}_t\in\R^{N\times D}_{+}$ via \eqref{eq:node_lift}. A macro inertial update and a micro KG-constrained geometric correction are then computed, fused through the geometric gate, and retracted to the positive cone before readout.

\noindent\textbf{Masking for visibility and supervision:}
The binary mask $\vM_{1:T}$ is not appended to the node features. Instead, it is used to hide unobserved nodes/entries by zero-filling the corresponding inputs and define $\Omega_{\mathrm{tar}}$ for supervision and evaluation.

\subsection{Input Representation and Positive-Cone Lifting}
\label{sec:signal_rep}

To capture temporal dynamics beyond instantaneous values, we embed each standardized measurement $\tilde{x}_{i,t}$ into a $K$-dimensional time-delay (sliding-window) vector:
\begin{equation}\label{eq:phase_embed}
\vx_{i,t} = [\tilde{x}_{i,t}, \tilde{x}_{i,t-\tau}, \dots, \tilde{x}_{i,t-(K-1)\tau}]^\top \in \R^{K},
\end{equation}
where $\tau$ is the delay and $K$ is the embedding dimension. Stacking all nodes yields a phase-space snapshot $\vX_t\in\R^{N\times K}$, and $\vX_{1:T}$ denotes the resulting sequence.

Masking and padding:
Under the systemic blind spot protocol in Definition~\ref{def:blind_spot}, missingness occurs as a node-wise block over $\Tb$, hence all lagged components of the masked nodes are unavailable. In implementation, after z-score normalization we replace unobserved samples by zeros (the normalized mean), and we use standard zero padding when $t-\ell\tau<1$. We do not concatenate $\vM_{1:T}$ as an additional feature; instead, it is used to zero-fill the corresponding inputs in $\vX_{1:T}$ and to define $\Omega_{\mathrm{tar}}$ for supervision and evaluation.

\noindent\textbf{Energetic rectification (positivity-preserving map):}
To support geometry-aware similarity, we map heterogeneous (possibly signed) measurement histories to a positive latent space via a smooth element-wise function; in implementation we employ Softplus:
\begin{equation}
\operatorname{Softplus}(z)=\ln(1+e^{z}),\qquad
\vh\leftarrow \operatorname{Softplus}(\vh).
\label{eq:softplus}
\end{equation}
Softplus yields strictly positive coordinates, which will also ensure the induced spherical representatives lie in the interior of $\Sphere^{D-1}_{+}$.

\noindent\textbf{Latent positive lifting:}
Raw wireless measurements are heterogeneous and can be signed (for instance, power in dBm), and they are typically computed as windowed statistics of random link and traffic processes. We view the lifting as mapping such heterogeneous observations into a latent intensity space. In RieIF we do not assume the observed values form a probability vector. Instead, we learn a lifting map that rectifies and reshapes each phase-space observation into a nonnegative latent coordinate $\vh_{i,t}\in\Mp\subseteq\R^{D}_{+}$.

\begin{definition}[Positive-cone lifting]
	\label{def:positive_measure}
	We employ a learnable node-wise lifting map $\Phi_{\text{node}}: \R^{K} \to \Mp$ and define $\vh_{i,t}=\Phi_{\text{node}}(\vx_{i,t})\in \Mp\subseteq \R^{D}_{+}$. This representation produces nonnegative latent coordinates, enabling a positivity-preserving chart and scale--shape disentanglement.
\end{definition}

\noindent\textbf{Positive-cone lifting (node-wise encoder):}
We instantiate $\Phi_{\text{node}}$ as a learnable projection followed by Softplus:
\begin{equation}
\vh^{(0)}_{i,t} = \Phi_{\text{node}}(\vx_{i,t})
= \operatorname{Softplus}\!\left(\vx_{i,t}\vW_{\text{in}}+\vb_{\text{in}}\right)
\in \R^{D}_{+},
\label{eq:node_lift}
\end{equation}
and stacking over $i$ yields $\vH^{(0)}_t\in \R^{N\times D}_{+}$.

\subsection{Macro Stream: Global Inertia}
\label{sec:temporal_stream}

This stream captures global inertia, including network-wide trends, providing a robust fallback when local neighborhoods are compromised by systemic blind spots.

\noindent\textbf{Global Context Projection:}
Let $\vX_t\in\R^{N\times K}$ denote the masked phase-space snapshot at time $t$ (Sec.~\ref{sec:signal_rep}). We form a global vector by flattening the tensor:
\begin{equation}
\vx^{\text{glob}}_t =
\operatorname{Flatten}(\vX_t)
\in \R^{N\cdot K}.
\end{equation}
We then project it into $\Mp$:
\begin{equation}
\vh^{\text{macro}}_t
=\operatorname{Softplus}\!\left(\vx^{\text{glob}}_t\vW_{\text{glob}}+\vb_{\text{glob}}\right)
\in \R^{D}_{+}.
\label{eq:holo_proj}
\end{equation}
Interpretation: This projection learns a global ``energy coordinate'' of the system: even when a subset of nodes has missing observations, the remaining observed structure still contributes to $\vh^{\text{macro}}_t$, yielding a stable inertial anchor.

\noindent\textbf{Global Dynamics in the Tangent Chart:}
We model global evolution with an LSTM network, interpreted as producing a tangent update (velocity) rather than a constrained state:
\begin{equation}
\begin{aligned}
\vu_{\text{macro},t},(\vct{c}_t,\vct{s}_t)
&= \operatorname{LSTM}\!\left(\vh^{\text{macro}}_t,(\vct{c}_{t-1},\vct{s}_{t-1})\right),\\
\vu_{\text{macro},t} &\in \R^{D}.
\end{aligned}
\label{eq:macro_vel}
\end{equation}
Legitimacy of negative outputs: Negative components are permissible because $\vu_{\text{macro},t}$ lives in the (local Euclidean) tangent chart rather than on $\Mp$.

\noindent\textbf{Systemic Broadcasting:}
We broadcast the global update to all nodes:
\begin{equation}
\mat{U}_{\text{time},t}=\operatorname{Broadcast}(\vu_{\text{macro},t})\in\R^{N\times D}.
\label{eq:broadcast}
\end{equation}

\subsection{Micro Stream: KG-Constrained Shape Correction}
\label{sec:spatial_stream}

The micro stream produces a local geometric correction guided by the invariant knowledge graph $\KG$ (Sec.~\ref{sec:kg_topology}). It is designed to remain effective when data-driven correlations become unreliable under systemic blind spots.

\noindent Following Sec.~\ref{sec:preliminaries}, the micro stream computes scale-invariant, Fisher--Rao-aligned similarities on the spherical chart for KG-constrained aggregation, while accumulating updates in the Euclidean (tangent) chart.

\noindent\textbf{Spherical chart for attention:}
We apply the mapping in~\eqref{eq:spherical_mapping} to the node-wise embeddings $\vh^{(0)}_{i,t}$ and obtain
\begin{equation}
\vz_{i,t}=\frac{\vh^{(0)}_{i,t}}{\|\vh^{(0)}_{i,t}\|_2+\epsilon}\in\Sphere^{D-1}_{+}.
\label{eq:micro_sphere}
\end{equation}
We use $\vz_{i,t}$ to compute Fisher--Rao-aligned angular similarities between the induced simplex points $\vct{p}_{i,t}=\vz_{i,t}\odot\vz_{i,t}$. In implementation, we additionally $\ell_2$-normalize the projected queries/keys (namely, apply \texttt{F.normalize} to $\vct{q}_{i,t}$ and $\vct{k}_{j,t}$) before the dot product in \eqref{eq:kg_attn}, so the attention score is a cosine similarity. By \eqref{eq:fisher_rao_dist}, this cosine score is a monotone surrogate of the Fisher--Rao distance.

\noindent\textbf{KG-masked multi-head attention:}
We perform masked multi-head attention over the KG, following the scaled dot-product attention used in Transformers~\cite{vaswani2017attention} and its graph variants~\cite{velickovic2017graph}.

Queries and keys incorporate Laplacian positional encoding as in~\eqref{eq:lap_pe} (cf.\ Sec.~\ref{sec:preliminaries}).

\noindent Protocol-consistent attention mask: The KG encodes protocol dependencies and can be directed. We use its adjacency as a hard mask and define $\NKG{i}:=\{j\mid (j,i)\in\Ekg\}$ so that information flows only along allowed dependency directions.

We compute attention only over the KG neighborhood $\NKG{i}$:
\begin{equation}
\alpha_{ij,t}=\operatorname{Softmax}_{j\in\NKG{i}}
\left(
\frac{\langle \vct{q}_{i,t},\ \vct{k}_{j,t}\rangle}{\sqrt{d_k}}
\right).
\label{eq:kg_attn}
\end{equation}
where $d_k$ is the query/key dimension (per head).
Key design choice: Queries/keys are computed from normalized $\vz$ to ensure scale-invariant scoring, but the values are computed from the unnormalized positive-cone embeddings to carry magnitude information:
\begin{equation}
\vu_{\text{micro},i,t}=\sum_{j\in\NKG{i}} \alpha_{ij,t}\,\big(\vh^{(0)}_{j,t}\vW_V\big)
\in \R^{D}.
\label{eq:micro_vel}
\end{equation}
Stacking over $i$ gives $\mat{U}_{\text{space},t}\in\R^{N\times D}$.

\noindent\textbf{Local retraction:}
The micro update is formed in the Euclidean chart and then retracted back to $\Mp$ (cf. \eqref{eq:softplus}):
\begin{equation}
\begin{aligned}
\vH^{\text{micro}}_t
&=\mathcal{R}\!\left(\vH^{(0)}_t,\mat{U}_{\text{space},t}\right)\\
&=\operatorname{Softplus}\!\left(\vH^{(0)}_t+\mat{U}_{\text{space},t}\right),\\
\vH^{\text{micro}}_t &\in \R^{N\times D}_{+}.
\end{aligned}
\label{eq:micro_retract}
\end{equation}
\noindent Why Softplus update: We use Softplus as a smooth positivity-preserving update surrogate. An exact exponential map on the cone is unnecessary for our chart-based updates and can be numerically fragile, while Softplus provides stable gradients and prevents invalid negative states.

\subsection{Adaptive Synthesis and Training}
\label{sec:synthesis}

\noindent\textbf{Geometric Gating:}
We fuse the macro and micro tangent updates with an adaptive gate implemented by a lightweight multi-layer perceptron (MLP). We compute:
\begin{equation}
\vct{g}_{i,t}
=
\sigma\!\left(
\operatorname{MLP}_{\text{gate}}
\big([\vu_{\text{micro},i,t}\,\|\,\vu_{\text{macro},t}]\big)
\right)\in(0,1)^D,
\end{equation}
and fuse updates channel-wise:
\begin{equation}
\vu^{\text{total}}_{i,t}
=
\vct{g}_{i,t}\odot \vu_{\text{micro},i,t}
+
(\mathbf{1}-\vct{g}_{i,t})\odot \vu_{\text{macro},t}.
\label{eq:fuse_vel}
\end{equation}

Defensive fallback: Under systemic blind spots, the gate can suppress unreliable local corrections and revert to the global inertial update.

\noindent\textbf{Retraction and latent prediction:}
We obtain the final latent prediction by retracting the fused update:
\begin{equation}
\begin{aligned}
\widehat{\vH}_t
&=\mathcal{R}\!\left(\vH^{(0)}_t,\mat{U}^{\text{total}}_t\right)\\
&=\operatorname{Softplus}\!\left(\vH^{(0)}_t+\mat{U}^{\text{total}}_t\right),\\
\widehat{\vH}_t &\in \R^{N\times D}_{+}.
\end{aligned}
\label{eq:final_retract}
\end{equation}
where $\mat{U}^{\text{total}}_t$ stacks $\vu^{\text{total}}_{i,t}$ over $i$.

\noindent\textbf{Readout:}
The readout maps latent states back to the (z-score normalized) measurement space:
\begin{equation}
\hat{x}_{i,t}=\vW_{\text{out}}\widehat{\vh}_{i,t}+\vb_{\text{out}},
\label{eq:readout}
\end{equation}
where the readout is linear to cover $\R$ (while internal representations remain in $\R^{D}_{+}$ for geometric stability).

\noindent\textbf{Training objective:}
RieIF is trained to predict only the masked entries, matching the systemic blind spot evaluation protocol. Let $\Omega_{\mathrm{tar}}:=\{(i,t)\mid \vM_{i,t}=1\}$ denote the masked index set used for supervision, and let $\hat{\vct{x}}_{\Omega_{\mathrm{tar}}}$ and $\vct{x}^{\text{gt}}_{\Omega_{\mathrm{tar}}}$ be the vectorized estimate and ground truth on $\Omega_{\mathrm{tar}}$. The model parameters are learned by minimizing
\begin{align}
\mathcal{L}
=&\;\lambda_{\text{scale}}\underbrace{\frac{1}{|\Omega_{\mathrm{tar}}|}\sum_{(i,t)\in\Omega_{\mathrm{tar}}}(\hat{x}_{i,t}-x^{\text{gt}}_{i,t})^2}_{\text{mean squared error (scale)}}\nonumber\\
&\;+\lambda_{\text{shape}}\underbrace{\Bigg(1-\frac{\langle \hat{\vct{x}}_{\Omega_{\mathrm{tar}}},\vct{x}^{\text{gt}}_{\Omega_{\mathrm{tar}}}\rangle}{\|\hat{\vct{x}}_{\Omega_{\mathrm{tar}}}\|_2\,\|\vct{x}^{\text{gt}}_{\Omega_{\mathrm{tar}}}\|_2+\epsilon}\Bigg)}_{\text{Cosine (shape)}}.
\label{eq:loss_final}
\end{align}

The mean squared error term enforces numerical accuracy (scale), while the cosine term encourages trend and shape alignment. AdamW with weight decay is adopted as implicit $\ell_2$ regularization, and no extra penalty terms are added.

\begin{algorithm}[t]
\caption{RieIF Training and Inference}
\label{alg:rieif}
\begin{algorithmic}[1]
\STATE \textbf{Input:} Masked segment $(\vX_{1:T},\vM_{1:T})$, knowledge graph $\KG$
\STATE \textbf{Output:} Predicted values $\hat{\vX}_{1:T}$
\STATE Phase-space inputs $\vx_{i,t}$ are constructed by time-delay embedding in~\eqref{eq:phase_embed} with zero-filling according to $\vM_{1:T}$
\STATE Node embeddings are lifted to the positive cone: $\vh^{(0)}_{i,t}\leftarrow \operatorname{Softplus}(\Phi_{\mathrm{in}}(\vx_{i,t}))$, and normalized as $\vz_{i,t}\leftarrow \vh^{(0)}_{i,t}/(\|\vh^{(0)}_{i,t}\|_2+\epsilon)$
\STATE Macro stream update is computed: $u^{\mathrm{macro}}_t\leftarrow \operatorname{LSTM}(\operatorname{Softplus}(\Phi_{\mathrm{glob}}(\operatorname{Flatten}(\vX_t))))$ and broadcast to all nodes
\STATE Micro stream update $u^{\mathrm{micro}}_{i,t}$ is produced by KG-masked attention using $Q/K$ from $\vz$ (with Laplacian positional encoding) and $V$ from $\vh^{(0)}$
\STATE Tangent updates are fused: $u^{\mathrm{total}}_{i,t}\leftarrow g_{i,t}\odot u^{\mathrm{micro}}_{i,t}+(1-g_{i,t})\odot u^{\mathrm{macro}}_{t}$
\STATE Positivity-preserving update is applied: $\hat{\vh}_{i,t}\leftarrow \operatorname{Softplus}(\vh^{(0)}_{i,t}+u^{\mathrm{total}}_{i,t})$
\STATE Readout is computed: $\hat{x}_{i,t}\leftarrow \Phi_{\mathrm{out}}(\hat{\vh}_{i,t})$
\STATE Parameters are learned by minimizing \eqref{eq:loss_final} over $(i,t)\in\Omega_{\mathrm{tar}}$ using Adam~\cite{Kingma2015Adam}
\end{algorithmic}
\end{algorithm}

\noindent Algorithm~\ref{alg:rieif} summarizes the overall training and inference procedure.

\subsection{Computational Complexity}
\label{sec:complexity}
Let $D$ denote the latent dimension, $|\Ekg|$ the number of knowledge-graph edges, and $H$ the number of attention heads (with $d_k=D/H$). Per layer and time step, the micro stream costs $\mathcal{O}(H|\Ekg|d_k)=\mathcal{O}(|\Ekg|D)$, while gating/retraction/readout are $\mathcal{O}(ND)$ and the macro LSTM is $\mathcal{O}(D^2)$. Hence the overall per-step complexity is $\mathcal{O}(|\Ekg|D+ND+D^2)$. Using the sparse KG avoids the $\mathcal{O}(N^2D)$ cost of fully-connected attention; representative operation counts are summarized in Table~\ref{tab:computational_cost}. Importantly, all geometric operations used at inference are closed form, involving $\ell_2$ normalization, masked dot products, and Softplus retraction, without iterative manifold optimization.

\begin{table}[t]
\centering
\caption{Computational Cost Comparison (Representative Setting)}
\label{tab:computational_cost}
\renewcommand{\arraystretch}{1.10}
\setlength{\tabcolsep}{4pt}
\resizebox{\columnwidth}{!}{
\begin{tabular}{l l r c}
\toprule
\textbf{Model} & \textbf{Dominant mixing term} & \textbf{Est. Ops} & \textbf{Rel.}\\
\midrule
\textbf{RieIF (Ours)} & $BT_{\mathrm{seg}}L\,|\Ekg|\,D$ & $\sim$3.3M & 1.0$\times$\\
GinAR~\cite{yu2024ginar} & $BT_{\mathrm{seg}}L\,N^2D$ & $\sim$76M & 23.1$\times$\\
SE-HTGNN~\cite{wang2025simple} & $BT_{\mathrm{seg}}L\,(|\cE|{+}N)D$ & $\sim$5.5M & 1.7$\times$\\
CoIFNet~\cite{tang2025coifnet} & $BT_{\mathrm{seg}}\,(N^2{+}|\cE|)D$ & $\sim$40M & 12.1$\times$\\
\bottomrule
\end{tabular}}
\begin{flushleft}
\footnotesize
Note: We report rough multiply--accumulate operation counts for the dominant variable-mixing operations per training segment using $B{=}16$, $T_{\mathrm{seg}}{=}32$, $N{=}34$, $D{=}64$, $L{=}2$, and $|\Ekg|\approx|\cE|\approx 50$. Shared MLP/readout costs are omitted, so the table is intended for relative comparison.
\end{flushleft}
\end{table}

\subsection{Analysis of Blind Spot Robustness}
\label{sec:theory}
This subsection provides a concise analysis of why systemic blind spots are intrinsically difficult and how the proposed design mitigates the resulting failure modes.

\noindent\textbf{Irreducible error under proxy--target masking.}
Let $\tilde{x}_{\istar,t}$ denote a masked target entry and let $z$ denote the remaining observations available to an estimator during a blind spot.
Under squared loss, the optimal estimator is $g^\star(z)=\mathbb{E}[\tilde{x}_{\istar,t}\mid z]$ and the minimum achievable risk equals the conditional variance:
\begin{equation}
\inf_{g}\ \mathbb{E}\bigl[(\tilde{x}_{\istar,t}-g(z))^2\bigr] \;=\; \mathbb{E}\!\left[\mathrm{Var}(\tilde{x}_{\istar,t}\mid z)\right].
\label{eq:bayes_risk_condvar}
\end{equation}
When the mask removes the strongest proxies in $\Pof{\istar}$, the conditioning set $z$ becomes weakly informative and $\mathrm{Var}(\tilde{x}_{\istar,t}\mid z)$ approaches the marginal variance, implying a large irreducible error.
A local Gaussian surrogate makes this explicit: if $(\tilde{x}_{\istar,t},z)$ is approximated as jointly Gaussian with cross-covariance $r$ and covariance $\Sigma$ for $z$, then $\mathrm{Var}(\tilde{x}_{\istar,t}\mid z)=\sigma_{\istar}^2-r^\top\Sigma^{-1}r$, and proxy masking reduces $\|r\|$, thereby increasing the Bayes risk in \eqref{eq:bayes_risk_condvar}.
\noindent\textbf{Geometry mismatch behind Euclidean ``tunneling''.}
Let $\vh_a$ and $\vh_b$ denote the latent states at the two ends of a blind spot, constrained to a curved feasible set induced by positivity and compositional structure.
Euclidean interpolation and dot-product aggregation connect $(\vh_a,\vh_b)$ through a straight chord in the ambient space, which can deviate substantially from a manifold-consistent path when the intrinsic curvature is non-negligible.
This mismatch becomes pronounced exactly when blind spots are long, because the endpoints are farther apart and local linearization is unreliable.

\noindent\textbf{Why RieIF is stable under blind spots.}
RieIF alleviates information collapse by (i) injecting a protocol-derived knowledge graph that remains available during outages, and (ii) performing scale-invariant aggregation on the positive unit hypersphere, where angular similarity aligns with Fisher--Rao geometry.
Moreover, the positivity-preserving retraction $\operatorname{Softplus}(\cdot)$ is non-expansive because its derivative lies in $(0,1)$, and the geometric gate forms a convex combination of macro and micro tangent updates. Together, these properties promote bounded transport and provide a robust fallback when local statistical evidence is insufficient.

\section{Experimental Evaluation}\label{sec:experiments}

\subsection{Experimental Setup}

\noindent\textbf{Wireless Datasets:} We evaluate on three wireless datasets collected across different network layers.

\paragraph{Network Monitoring (network-level throughput)} This dataset contains full-stack 5G/6G KPIs collected at a deployed network for throughput prediction. From 86 raw KPIs, we select $N{=}34$ protocol-aligned data-field nodes and use the protocol-derived knowledge graph in Sec.~\ref{sec:kg_topology}. The target is the uplink throughput (for instance, dual-connectivity PHY throughput), whose strongest proxies include the cross-layer throughput and link-adaptation statistics such as MCS, PRB usage, and BLER.

\paragraph{Beam Prediction (system-level EVM)}  This is a 6G Integrated Sensing and Communication (ISAC) dataset for UAV beam tracking. The data is collected from a mmWave communication testbed and comprises beamspace EVM measurements and UAV-sensed kinematic data. The target is the average EVM, and we build a lightweight physical-dependency knowledge graph.

\paragraph{Link Adaptation (link-level post-SINR)} The dataset is generated by a 5G New Radio (NR) link-adaptation simulator, for MCS selection. The inputs include CQI statistics, RI, MCS, ACK/NACK, BLER, and throughput; the target is the post-equalization SINR (dB). A control-loop KG is derived from the link-adaptation pipeline.

\noindent\textbf{Baseline Models:}
To ensure a comprehensive evaluation, we compare RieIF against three distinct categories of baselines, spanning classical temporal extrapolation, Euclidean spatio-temporal graph learning, and recent missing-data-aware architectures.
\setcounter{paragraph}{0}
\paragraph{Category I: Non-Deep Learning Baselines}
We include linear interpolation, spline interpolation~\cite{deBoor1978Splines}, TRMF~\cite{yu2016temporal}, and Kalman filtering~\cite{Kalman1960}. Additionally, we evaluated 17 traditional estimators; for brevity, Table~\ref{tab:overall_performance_snr} reports the best-performing one per dataset as Non-deep-learning (Best).

Role: These methods rely on temporal continuity or global low-rank assumptions. Under systemic blind spots, they often collapse to over-smoothed or flatline estimates, providing a clean control group for demonstrating the tunneling failure mode.

\paragraph{Category II: Classical Euclidean ST-GNNs}
We compare against established spatio-temporal graph networks including STGCN~\cite{yu2017spatio}, ASTGCN~\cite{guo2019attention}, GraphWaveNet~\cite{wu2019graph}, and STG2Seq~\cite{bai2019stg2seq}. These models incorporate spatial inductive bias but operate strictly in Euclidean feature space, enabling isolation of the benefit of geometry-consistent flow. For fairness, all graph-based baselines use the same protocol-derived KG topology.

\paragraph{Category III: Recent Advances (2024--2025)}
We include recent missing-data/graph learning frameworks: \textbf{GinAR}~\cite{yu2024ginar}, \textbf{SE-HTGNN}~\cite{wang2025simple} (efficient heterogeneous temporal graph neural network), and \textbf{CoIFNet}~\cite{tang2025coifnet}.

These baselines jointly cover temporal smoothing and low-rank priors, Euclidean spatio-temporal graph modeling, and recent missingness-aware designs, providing a balanced comparison spectrum. 

\noindent\textbf{Evaluation Metrics:}
We report standard regression metrics: Mean Absolute Error (MAE), Mean Squared Error (MSE), Root Mean Squared Error (RMSE), and the Coefficient of Determination ($R^2$). Unless otherwise stated, all metrics are computed on the masked indices $\Omega_{\mathrm{tar}}:=\{(i,t)\,|\,\vM_{i,t}=1\}$.

Signal-fidelity metric: To quantify fidelity in an energy sense, we report the recovery signal-to-noise ratio (SNR) in dB:
\begin{equation}
\begin{aligned}
\mathrm{SNR}(\mathrm{dB})
&=10\log_{10}
\frac{\sum_{(i,t)\in\Omega_{\mathrm{tar}}} (x^{\mathrm{gt}}_{i,t})^2}{\sum_{(i,t)\in\Omega_{\mathrm{tar}}} (x^{\mathrm{gt}}_{i,t}-\hat{x}_{i,t})^2}.
\end{aligned}
\end{equation}
A $+3$\,dB increase corresponds to approximately doubling the fidelity in an energy ratio, and SNR provides an intuitive robustness indicator under systemic blind spots.

\noindent\textbf{Implementation details:}
We adopt an 80/20 chronological split without timestamp shuffling; a validation subset is carved out from the training portion for early stopping. Unless otherwise stated, each sample is a segment of length $T_{\mathrm{seg}}=32$ with time-delay embedding dimension $K=5$ (Sec.~\ref{sec:signal_rep}); the blind spot block length is set to $|\Tb|=32$ in the default setting. To stress-test geometry under structured information collapse, we adopt the systemic blind spot protocol consistent with Definition~\ref{def:blind_spot} in Sec.~\ref{sec:problem_formulation}. For each trial, we compute Pearson correlations on the training split to form the proxy set $\Pof{\istar}=\{j:\,|\mathrm{corr}(\tilde{x}_{\istar,\cdot},\tilde{x}_{j,\cdot})|\ge\rho\}$. We then sample a contiguous block $\Tb$ and mask the node set $\{\istar\}\cup \Pof{\istar}$ within $\Tb$, yielding the binary mask $\vM_{1:T}$. The same rule is used to generate training/validation and test masks, always recomputing correlations on the corresponding training split to prevent information leakage.

All deep models are trained with AdamW (weight decay $10^{-5}$) for up to 100 epochs using cosine learning-rate decay and early stopping (patience 15); gradients are clipped for stability. RieIF uses latent dimension $D=32$, attention heads $H=8$, graph-transformer layers $L=2$, a one-layer LSTM macro stream, Laplacian positional encoding dimension 16, and batch size 32. The learning rate is selected via validation search per dataset. RieIF is trained with the hybrid loss in \eqref{eq:loss_final}; the ablation No Geo. Loss removes the cosine term.

\subsection{Main Performance Analysis}

\noindent\textbf{Empirical manifold evidence:}
The discussion in Sec.~\ref{sec:preliminaries} motivates a geometry-consistent prediction view under structured missingness, where Euclidean interpolation can tunnel across missing blocks on a curved constraint manifold. Here we provide dataset-level evidence that the wireless data state space is non-flat and that Euclidean distances systematically underestimate intrinsic transport distances.

\paragraph{Isomap-style geodesic approximation}
We uniformly sample $T_{\mathrm{s}}=\min(T,2000)$ wireless data snapshots from the training split, build a symmetrized $k_{\mathrm{nn}}$-nearest-neighbor graph ($k_{\mathrm{nn}}=15$) in the normalized data space with edge weights $\|\tilde{\vct{x}}_{t_a}-\tilde{\vct{x}}_{t_b}\|_2$, and run Dijkstra shortest paths to obtain a graph-geodesic surrogate $d_{\text{geo}}(t_a,t_b)$. We compare it with the Euclidean distance $d_{\text{Euc}}(t_a,t_b)=\|\tilde{\vct{x}}_{t_a}-\tilde{\vct{x}}_{t_b}\|_2$ and report the distortion ratio $d_{\text{geo}}/d_{\text{Euc}}$ and violation rate $\Pr[d_{\text{geo}}>d_{\text{Euc}}]$.

\begin{figure*}[!t]
    \centering
    \includegraphics[width=0.95\textwidth,trim=2mm 2mm 2mm 2mm,clip]{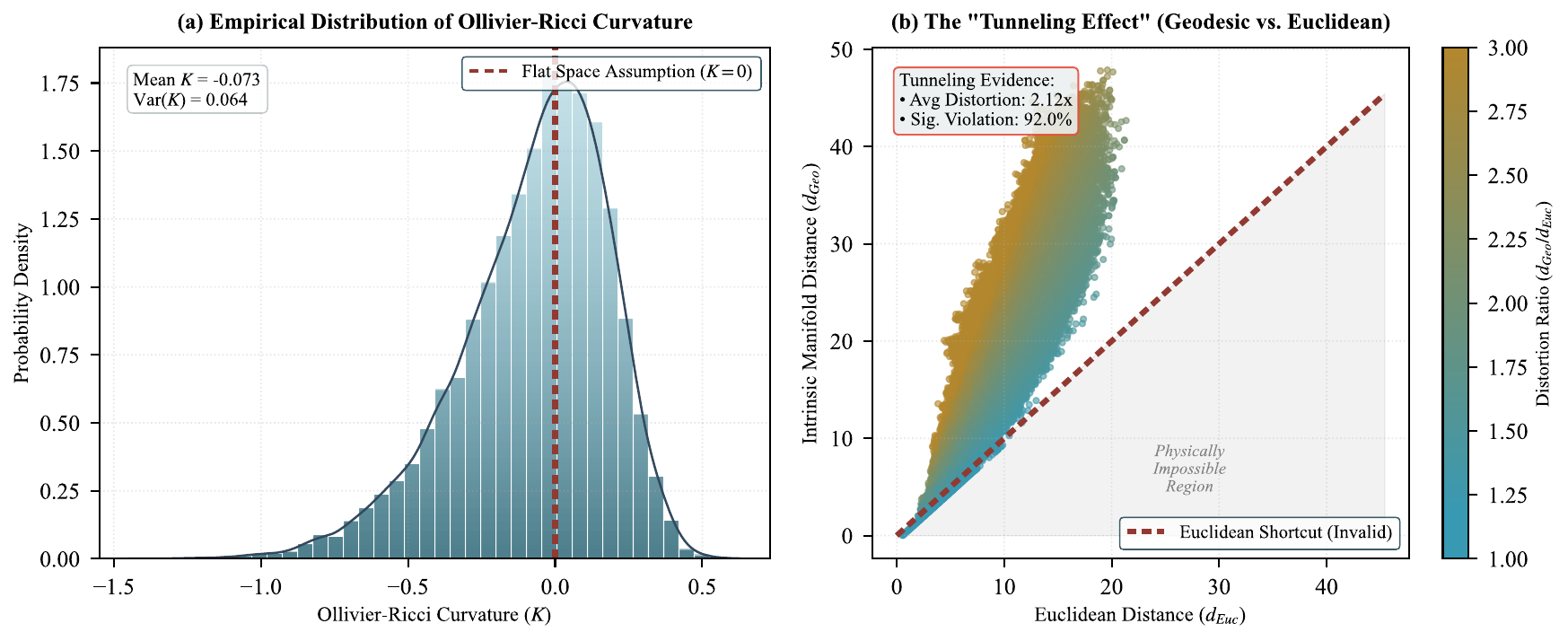}
    \caption{Non-Euclidean evidence in wireless data state space. (a) Ollivier--Ricci curvature on the $k_{\mathrm{nn}}$-nearest-neighbor graph ($k_{\mathrm{nn}}=15$). (b) Euclidean versus graph-geodesic distances; distortion $d_{\mathrm{geo}}/d_{\mathrm{Euc}}>1$ indicates Euclidean tunneling.}
    \label{fig:manifold_evidence}
\end{figure*}

\paragraph{Curvature and distortion diagnostics}
Fig.~\ref{fig:manifold_evidence} summarizes two complementary diagnostics computed on the Network Monitoring dataset:
(i) the empirical distribution of Ollivier--Ricci curvature on the neighborhood graph (non-zero mean/variance, hence non-flat), and
(ii) a direct geodesic-vs-Euclidean distortion plot. In Fig.~\ref{fig:manifold_evidence}(b), the average distortion is 2.12$\times$ and 92.0\% of sampled pairs satisfy $d_{\text{geo}}>d_{\text{Euc}}$, indicating pervasive Euclidean underestimation.

With these manifold diagnostics in place, we next summarize the prediction performance of RieIF and baselines in Table~\ref{tab:overall_performance_snr}. We report standard error metrics alongside recovery SNR (dB) under the systemic blind spot protocol.

\noindent\textbf{Why classical baselines fail:}
As reflected by the Non-deep-learning (Best) rows in Table~\ref{tab:overall_performance_snr}, classical temporal and statistical estimators can be brittle under systemic blind spots.
Masking a target together with its strongest proxies removes the local evidence that such methods rely on, which often leads to unstable fits and low SNR.
This behavior is consistent with the non-identifiability and observability collapse discussed in Sec.~\ref{sec:theory}.

\noindent\textbf{Network Monitoring results:}
RieIF achieves the largest relative improvement on the Network Monitoring dataset under the default blind spot setting ($\rho=0.4$, $|\Tb|=32$).
It attains $R^2=0.9483$ and SNR$=12.87$\,dB, improving over the strongest baseline (CoIFNet, 9.64\,dB) by 3.23\,dB.
In terms of RMSE, this corresponds to a drop from 0.3165 to 0.2182, that is, a 31.1\% relative reduction.
This gain is consistent with the manifold diagnostics in Fig.~\ref{fig:manifold_evidence} and supports geometry-consistent flow under systemic blind spots (Fig.~\ref{fig:euclidean_fallacy}).

\noindent\textbf{Beam Prediction results:}
On the Beam Prediction dataset, RieIF attains $R^2=0.9333$ and SNR$=11.76$\,dB, improving over the strongest baseline (CoIFNet, 11.08\,dB) by 0.68\,dB.
This setting is driven by UAV-induced non-stationarity and beam/attitude dynamics, and the gain indicates that the proposed geometry-aware transport remains effective beyond the cellular data-field graph.

\noindent\textbf{Link Adaptation results:}
On the Link Adaptation dataset, RieIF achieves $R^2=0.9712$ and SNR$=15.41$\,dB, outperforming the strongest baseline (GinAR, 12.35\,dB) by 3.05\,dB.
In terms of RMSE, this corresponds to a drop from 0.2649 to 0.1864, that is, a 29.6\% relative reduction.
These results suggest that RieIF remains effective when the cross-layer wireless measurements are generated by a closed-loop link adaptation pipeline with rapidly varying channel conditions.

\noindent\textbf{Overall average:}
Averaged over the three wireless datasets in Table~\ref{tab:overall_performance_snr}, RieIF achieves $\overline{R^2}=0.9509$ and $\overline{\mathrm{SNR}}=13.34$\,dB, exceeding the best baseline (GinAR, 10.71\,dB) by $+2.63$\,dB in average SNR.

\begin{table*}[!t]
\centering
\caption{PREDICTION PERFORMANCE ACROSS WIRELESS DATASETS (mean over three seeds)}
\label{tab:overall_performance_snr}
\renewcommand{\arraystretch}{1.2}

\resizebox{\textwidth}{!}{
\begin{tabular}{l ccccc c ccccc}
\toprule[1.5pt]
\multirow{2.5}{*}{\textbf{Method}} & \multicolumn{5}{c}{\textbf{Network Monitoring (Part I)}} & & \multicolumn{5}{c}{\textbf{Beam Prediction (Part II)}} \\
\cmidrule(lr){2-6} \cmidrule(lr){8-12}
& $R^2 \uparrow$ & MSE $\downarrow$ & MAE $\downarrow$ & RMSE $\downarrow$ & SNR $\uparrow$ & & $R^2 \uparrow$ & MSE $\downarrow$ & MAE $\downarrow$ & RMSE $\downarrow$ & SNR $\uparrow$ \\
\midrule

Non-deep-learning (Best) & $-21.7692$ & 0.8061 & 0.8501 & 0.8978 & 0.58 & & $-0.0018$ & 1.1896 & 0.4827 & 1.0907 & $-2.09$ \\

STGCN & 0.5350 & 0.4284 & 0.5108 & 0.6545 & 3.33 & & 0.3707 & 0.4540 & 0.3336 & 0.6738 & 2.10 \\

ASTGCN & 0.4707 & 0.4877 & 0.6009 & 0.6983 & 2.76 & & 0.0089 & 0.7149 & 0.4086 & 0.8455 & 0.13 \\

GraphWaveNet & 0.7822 & 0.2007 & 0.3455 & 0.4480 & 6.62 & & 0.5371 & 0.3339 & 0.3018 & 0.5778 & 3.43 \\

STG2Seq & 0.7654 & 0.2161 & 0.3448 & 0.4649 & 6.30 & & 0.7403 & 0.1873 & 0.2384 & 0.4328 & 5.94 \\

SE-HTGNN & 0.8730 & 0.1170 & 0.2692 & 0.3421 & 8.96 & & 0.8791 & 0.0889 & 0.2061 & 0.2982 & 9.18 \\

GinAR & 0.8852 & 0.1057 & 0.2407 & 0.3252 & 9.40 & & 0.9086 & 0.0673 & \underline{0.1768} & 0.2594 & 10.39 \\

CoIFNet & \underline{0.8913} & \underline{0.1002} & \underline{0.2309} & \underline{0.3165} & \underline{9.64} & & \underline{0.9220} & \underline{0.0574} & 0.1807 & \underline{0.2396} & \underline{11.08} \\

\addlinespace[1pt]

\rowcolor[gray]{0.95}
\textbf{RieIF (Ours)} & \textbf{0.9483} & \textbf{0.0476} & \textbf{0.1536} & \textbf{0.2182} & \textbf{12.87} & & \textbf{0.9333} & \textbf{0.0491} & \textbf{0.1663} & \textbf{0.2216} & \textbf{11.76} \\

\bottomrule[1.5pt]
\end{tabular}
}

\vspace{4pt}

\resizebox{\textwidth}{!}{
\begin{tabular}{l ccccc c ccccc}
\toprule[1.5pt]
\multirow{2.5}{*}{\textbf{Method}} & \multicolumn{5}{c}{\textbf{Link Adaptation (Part III)}} & & \multicolumn{5}{c}{\textbf{Average (3 datasets)}} \\
\cmidrule(lr){2-6} \cmidrule(lr){8-12}
& $R^2 \uparrow$ & MSE $\downarrow$ & MAE $\downarrow$ & RMSE $\downarrow$ & SNR $\uparrow$ & & $R^2 \uparrow$ & MSE $\downarrow$ & MAE $\downarrow$ & RMSE $\downarrow$ & SNR $\uparrow$ \\
\midrule

Non-deep-learning (Best) & 0.1064 & 1.1786 & 0.8762 & 1.0856 & 0.10 & & $-7.2215$ & 1.0581 & 0.7363 & 1.0247 & $-0.47$ \\

STGCN & $-0.1990$ & 1.4458 & 0.9619 & 1.2024 & $-0.79$ & & 0.2356 & 0.7760 & 0.6021 & 0.8436 & 1.55 \\

ASTGCN & 0.2177 & 0.9434 & 0.7770 & 0.9713 & 1.07 & & 0.2324 & 0.7154 & 0.5955 & 0.8384 & 1.32 \\

GraphWaveNet & 0.8125 & 0.2261 & 0.3804 & 0.4755 & 7.27 & & 0.7106 & 0.2536 & 0.3426 & 0.5004 & 5.77 \\

STG2Seq & 0.4709 & 0.6381 & 0.6390 & 0.7988 & 2.76 & & 0.6589 & 0.3472 & 0.4074 & 0.5655 & 5.00 \\

SE-HTGNN & 0.5226 & 0.5758 & 0.6070 & 0.7588 & 3.21 & & 0.7583 & 0.2606 & 0.3608 & 0.4664 & 7.12 \\

GinAR & \underline{0.9418} & \underline{0.0702} & \underline{0.2119} & \underline{0.2649} & \underline{12.35} & & \underline{0.9119} & \underline{0.0811} & \underline{0.2098} & \underline{0.2832} & \underline{10.71} \\

CoIFNet & 0.9027 & 0.1173 & 0.2740 & 0.3425 & 10.12 & & 0.9053 & 0.0916 & 0.2285 & 0.2995 & 10.28 \\

\addlinespace[1pt]

\rowcolor[gray]{0.95}
\textbf{RieIF (Ours)} & \textbf{0.9712} & \textbf{0.0347} & \textbf{0.0899} & \textbf{0.1864} & \textbf{15.41} & & \textbf{0.9509} & \textbf{0.0438} & \textbf{0.1366} & \textbf{0.2087} & \textbf{13.34} \\

\bottomrule[1.5pt]
\end{tabular}
}
\begin{flushleft}
\footnotesize
Note: All values are the mean over three random seeds. Best results are highlighted in \textbf{bold}, and the second best are \underline{underlined}. SNR (dB) indicates recovery signal-to-noise ratio.
\end{flushleft}
\end{table*}

\subsection{Ablation Studies}
To validate the design rationale of RieIF, we conduct ablations on the Network Monitoring dataset under systemic blind spots. We remove or replace key components to isolate the contributions of geometry, topology, and architectural disentanglement. Results are summarized in Table~\ref{tab:ablation_study} using SNR drop (dB) relative to the full model.

\noindent\textbf{Geometry ablation:}
We replace Fisher--Rao (spherical) attention with Euclidean dot-product attention (Euclidean Attention) and remove the geometric supervision (No Geo. Loss, MSE only). In Euclidean Attention, we drop the normalization in \eqref{eq:micro_sphere} and compute queries/keys from $\vh^{(0)}$ directly, so attention becomes amplitude-sensitive.
Euclidean attention incurs a 2.02\,dB SNR drop, and removing geometric loss yields a 2.11\,dB drop. These results indicate that under magnitude volatility (such as fading or power control), Euclidean similarity is amplitude-biased, whereas spherical/Fisher--Rao-style alignment preserves shape consistency; the observed drops confirm that geometry materially improves blind spot recovery fidelity.

\noindent\textbf{Topology ablation:}
We compare the 3GPP-based KG against a data-driven correlation graph, a fully connected graph, and a random graph.
The KG achieves the best SNR (12.87\,dB), outperforming the correlation graph by 1.55\,dB and the fully connected alternative by 0.31\,dB. This is consistent with the blind spot setting: data-driven correlations become unreliable, whereas the protocol-derived KG remains a stable dependency backbone. Meanwhile, dense connectivity can inject global noise, supporting the selectivity-over-density premise.

\noindent\textbf{Macro--micro ablation:}
We remove either the Macro (temporal) stream or the Micro (spatial) stream.
Removing the Macro stream causes the largest degradation (5.04\,dB SNR drop), while removing the Micro stream yields a 2.47\,dB drop.
The macro stream provides the dead-reckoning backbone that carries the estimate through long blind spot intervals, whereas the micro stream supplies geometry-consistent corrections that reduce drift and restore physically plausible local shapes under KG constraints; both are needed to realize the full SNR gain.

\noindent\textbf{Gating mechanism:}
Global context projection: Replacing the global context projection (Linear $(N\cdot K)\to D$) with a node-wise lifting (Linear $K\to D$) incurs a 4.86\,dB drop, showing that system-level context is indispensable when a node's local history is masked.

Adaptive gating: Replacing the learnable gate with fixed fusion ($0.5/0.5$) causes a 1.47\,dB SNR loss, indicating that adaptive arbitration between macro and micro streams improves precision under structured uncertainty. The gate can be viewed as a reliability controller: when neighborhood evidence collapses inside a blind spot, it leans on the macro inertial update, while using micro corrections when KG messages remain informative.

\begin{table}[!t]
\centering
\caption{Ablation Study: Disentangling Geometric, Topological, and Architectural Contributions}
\label{tab:ablation_study}
\renewcommand{\arraystretch}{1.15}
\setlength{\tabcolsep}{3.5pt}

\resizebox{\columnwidth}{!}{
\begin{tabular}{l ccc c}
\toprule[1.5pt]
Model Variant & $R^2 \uparrow$ & RMSE $\downarrow$ & SNR (dB) $\uparrow$ & Drop (dB) \\
\midrule

\rowcolor[gray]{0.95}
\textbf{RieIF (Full Model)} & \textbf{0.9483} & \textbf{0.2182} & \textbf{12.87} & \textbf{--} \\
\addlinespace[2pt]

\multicolumn{5}{l}{Impact of Geometry} \\
\hspace{3mm} Euclidean Attention & 0.9177 & 0.2753 & 10.85 & $\downarrow$ 2.02 \\
\hspace{3mm} No Geo. Loss (MSE) & 0.9161 & 0.2781 & 10.76 & $\downarrow$ 2.11 \\
\addlinespace[2pt]

\multicolumn{5}{l}{Impact of Topology} \\
\hspace{3mm} Data-Driven Graph & 0.9262 & 0.2607 & 11.32 & $\downarrow$ 1.55 \\
\hspace{3mm} Fully Connected & 0.9446 & 0.2254 & 12.56 & $\downarrow$ 0.31 \\
\hspace{3mm} Random Graph & 0.8901 & 0.3183 & 9.59 & $\downarrow$ 3.28 \\
\addlinespace[2pt]

\multicolumn{5}{l}{Impact of Micro-Mechanism} \\
\hspace{3mm} No Adaptive Gate & 0.9276 & 0.2643 & 11.40 & $\downarrow$ 1.47 \\
\hspace{3mm} Node-wise Projection & 0.8417 & 0.3908 & 8.00 & $\downarrow$ 4.86 \\
\addlinespace[2pt]

\multicolumn{5}{l}{Impact of Dual-Stream (Macro--Micro)} \\
\hspace{3mm} No Macro Stream & 0.8350 & 0.3990 & 7.82 & $\downarrow$ 5.04 \\
\hspace{3mm} No Micro Stream & 0.9086 & 0.2901 & 10.39 & $\downarrow$ 2.47 \\

\bottomrule[1.5pt]
\end{tabular}
}

\begin{flushleft}
\vspace{2pt}
\scriptsize
Note: ``Drop'' indicates the SNR loss relative to the full model. ``Node-wise Projection'' refers to replacing the global context projection with a local history-only mapping.
\end{flushleft}
\end{table}

\begin{figure*}[!t]
	\centering
	\subfloat[Proxy-correlation sweep (SNR)]{
		\includegraphics[width=0.48\linewidth,trim=2mm 0mm 2mm 0mm,clip]{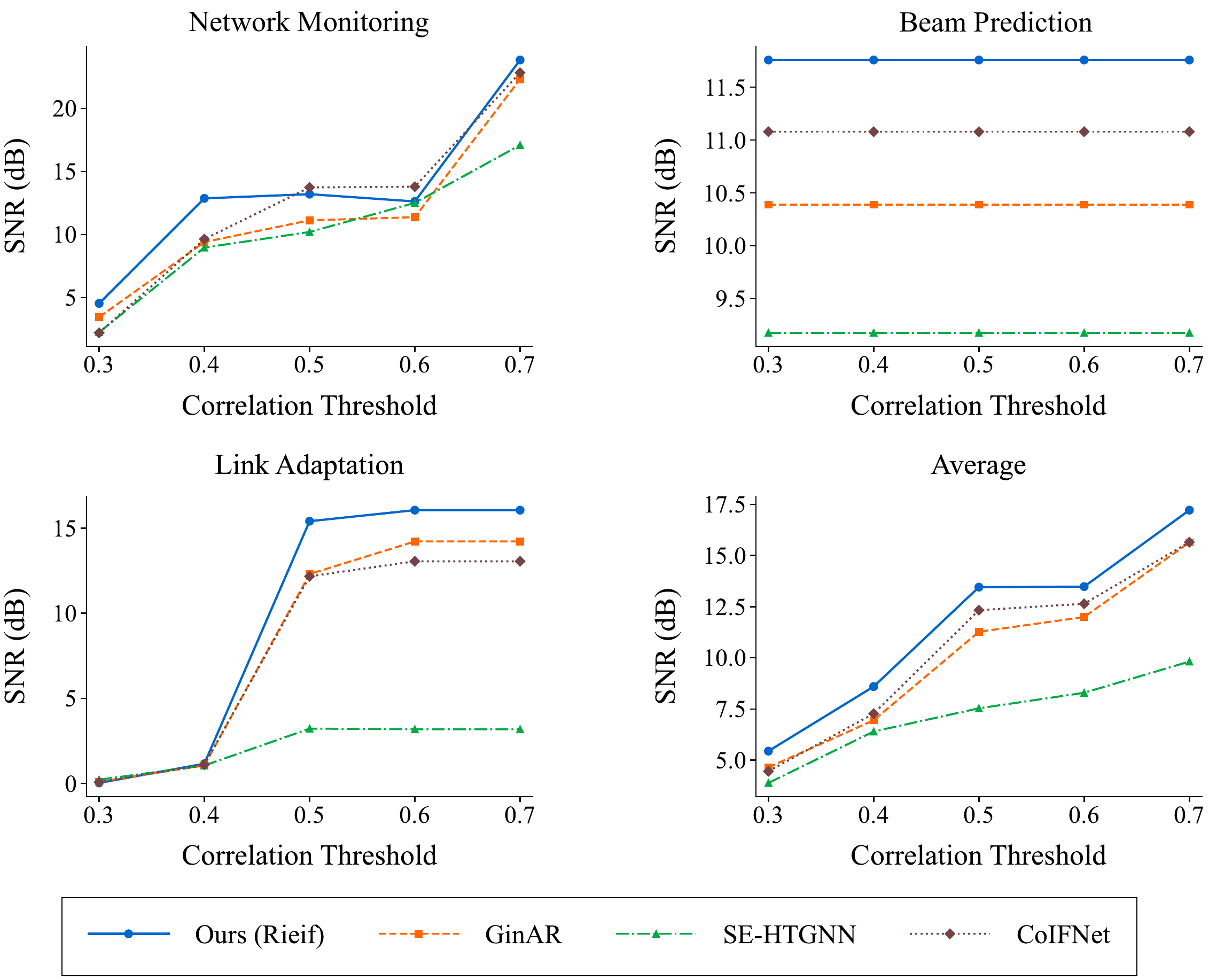}
	}
	\hfil
	\subfloat[Input-noise sweep (SNR)]{
		\includegraphics[width=0.48\linewidth,trim=2mm 0mm 2mm 0mm,clip]{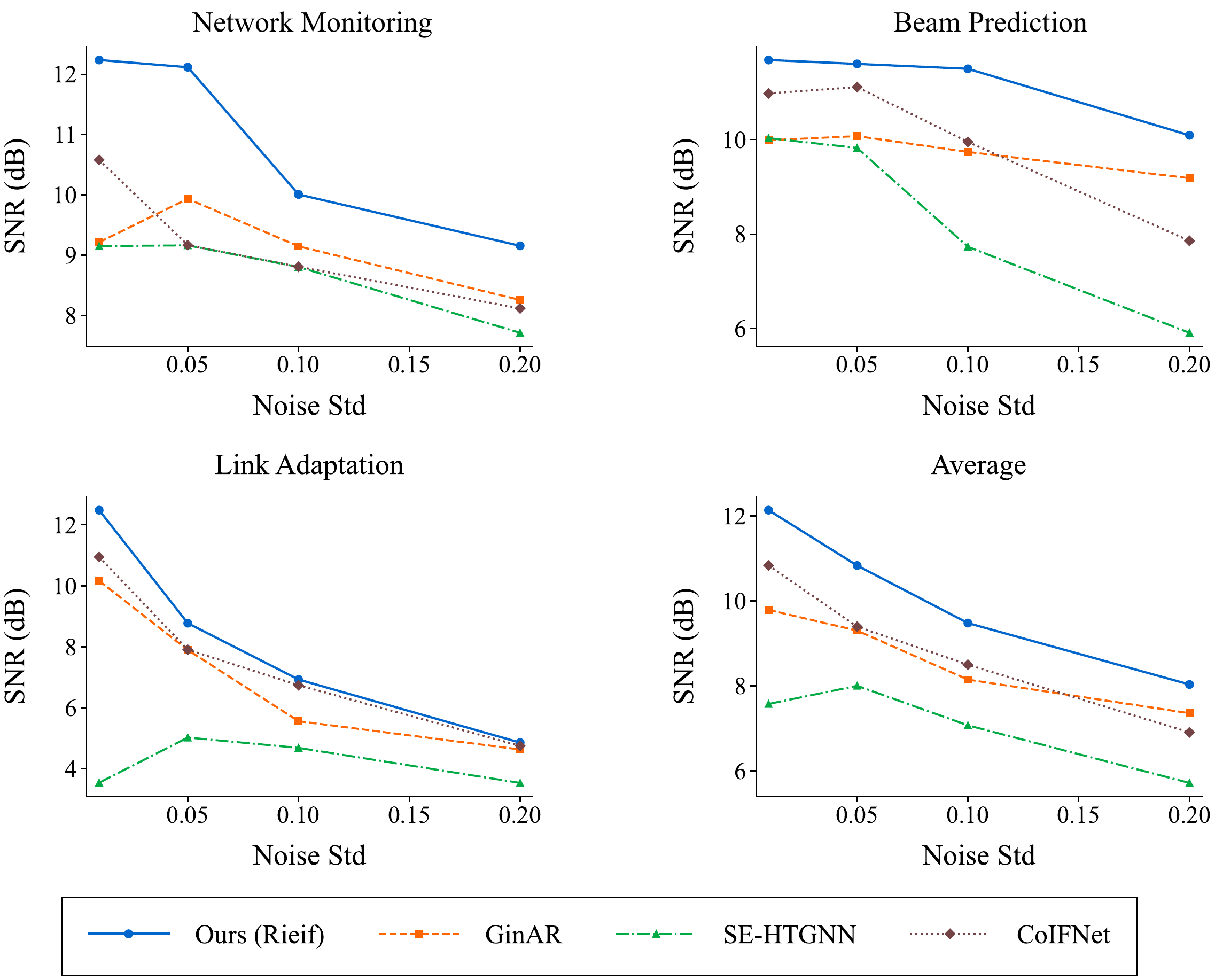}
	}
	\caption{Robustness under systemic blind spots (SNR). (a) Proxy-correlation threshold $\rho$ sweep (Network Monitoring and Link Adaptation; Beam Prediction excluded). (b) Input-noise sweep with Additive White Gaussian Noise (AWGN), $\sigma\in[0.01,0.20]$.}
	\label{fig:robustness_combined}
\end{figure*}

\subsection{Robustness Analysis}
We conduct robustness tests along (i) blind spot correlation (proxy selection threshold) and (ii) input noise, to verify that RieIF behaves as a stable geometric operator rather than overfitting to easy regimes.

\noindent\textbf{Correlation sensitivity:}
We vary the Pearson threshold $\rho$ used to define proxy sets $\Pof{\istar}$ in the systemic blind spot protocol. Lower $\rho$ yields weaker statistical neighborhoods and more severe information collapse.

The Beam Prediction dataset is excluded from this threshold sweep because proxy sets remain identical for $\rho\in[0.3,0.7]$: the target average EVM is almost perfectly correlated ($>0.98$) with two polarization-dependent EVM features, while correlations with all other features are below $0.3$.

Observation:
As illustrated in Fig.~\ref{fig:robustness_combined}(a), the proxy-correlation threshold $\rho$ controls the severity of systemic blind spots by expanding or shrinking the masked proxy set.
As $\rho$ decreases, more proxies are masked together with the target, correlation routes collapse, and all methods degrade.
RieIF remains consistently competitive because it constrains information flow with an invariant KG and uses scale-insensitive spherical interactions that better preserve directional structure when magnitudes drift.

Network Monitoring: at $\rho{=}0.4$, RieIF maintains 12.87\,dB SNR ($R^2{=}0.948$), exceeding CoIFNet by +3.23\,dB.
Link Adaptation: at $\rho{=}0.5$, RieIF achieves 15.41\,dB SNR ($R^2{=}0.971$), outperforming GinAR by about +3.1\,dB.
At very low thresholds on Link Adaptation ($\rho\le 0.4$), the proxy sets become overly broad and all methods experience a sharp performance drop, highlighting the difficulty of extreme information collapse.
The coefficient of determination follows the same trend and is omitted for brevity.

\noindent\textbf{Noise robustness:}
We inject AWGN into inputs, varying $\sigma$ from 0.01 to 0.20 (Fig.~\ref{fig:robustness_combined}(b)).

Observation 1: Non-monotonic behavior in Euclidean baselines.
Several Euclidean baselines exhibit mild-noise peaks (a stochastic-regularization effect) before degrading sharply at higher $\sigma$.

Observation 2: Geometric invariance of RieIF.
RieIF degrades gracefully and remains competitive even under severe noise, consistent with shape-focused transport on spherical charts (Sec.~\ref{sec:preliminaries}) rather than amplitude-sensitive similarity.

\section{Conclusion}\label{sec:conclusion}

This paper studied spatio-temporal graph signal prediction under structured missingness and noisy measurements in wireless networks. We focused on systemic blind spots, which leads to severe evidence collapse and exposes an inductive-bias mismatch in Euclidean interpolation and message passing. To tackle this challenge, we proposed RieIF, a knowledge-driven, geometry-consistent information-flow framework. For analytical tractability within the Fisher--Rao geometry, we projected the input from a Riemannian manifold onto a positive unit hypersphere, where angular similarity is computationally efficient. The KG was utilized as structural priors to constrain the attention mechanism in graph transformer and produced a micro stream. Meanwhile, an LSTM network modeled network temporal inertia and generated a macro stream. Finally, the two streams were adaptively fused for signal recovery, respectively emphasizing geometric shape and signal strength.

Experimental results on three wireless datasets consistently demonstrate performance gains under systemic blind spots. Specifically, on the Network Monitoring dataset, RieIF improves the recovery SNR from 9.64\,dB (CoIFNet) to 12.87\,dB and reduces the RMSE from 0.3165 to 0.2182. On the Link Adaptation dataset, it elevates the SNR from 12.35\,dB (GinAR) to 15.41\,dB while lowering the RMSE from 0.2649 to 0.1864. Ablation and robustness studies further corroborate that both the Fisher-Rao-aligned geometry and the protocol-derived knowledge graph are essential, enabling stable recovery even when correlation routes collapse or observation noise intensifies.

Future work will explore automated knowledge graph extraction from protocol documents and extend geometry-consistent transport to online diagnosis, forecasting, and closed-loop control.

\bibliographystyle{IEEEtran}
\bibliography{refs}

\end{document}